\documentclass[11pt,a4paper]{article}
\pdfoutput=1
\usepackage[utf8]{inputenc}
\usepackage{bbm}

\usepackage{fixmath}
\usepackage{jcappub}
\newcommand{\mathe}{\mathrm{e}}
\newcommand{\mathi}{\mathrm{i}}
\newcommand{\mathpi}{\pi}
\let\oldre\Re
\let\oldim\Im
\renewcommand{\Re}{\oldre\mathfrak{e}\,}
\renewcommand{\Im}{\oldim\mathfrak{m}\,}

\newcommand{\dalembert}{\square}
\newcommand{\total}{\operatorname{d}\!}
\newcommand{\christoffel}[4][]{\,{}{#1}\Gamma^{#2}_{#3 #4}\,}
\newcommand{\varchristoffel}[4][]{\,{}{#1}\tilde\Gamma^{#2}_{#3 #4}\,}
\newcommand{\eqend}[1]{\,\text{#1}}
\newcommand{\bigo}[1]{\mathcal{O}\left({#1}\right)}

\newcommand{\Ein}{\operatorname{Ein}}
\newcommand{\Li}{\operatorname{Li}}

\DeclareMathOperator*{\distlim}{d-lim}

\newcommand{\tr}{\operatorname{tr}}
\newcommand{\abs}[1]{{\left\lvert{#1}\right\rvert}}
\newcommand{\bra}[1]{{\left\langle{#1}\right\rvert}}
\newcommand{\ket}[1]{{\left\lvert{#1}\right\rangle}}
\newcommand{\expect}[1]{{\left\langle{#1}\right\rangle}_\text{c}}

\newcommand{\op}[1]{\hat{#1}}
\newcommand{\unitmatrix}{\mathbbm{1}}
\renewcommand{\vec}[1]{\mathbold{#1}}
\newcommand{\laplace}{\triangle}

\newcommand{\sgneps}{\epsilon^\pm}

\title{Riemann correlator in de Sitter including loop corrections from conformal fields}
\author[a]{Markus B. Fröb,}
\author[b]{Albert Roura}
\author[a]{and Enric Verdaguer}
\affiliation[a]{Departament de Física Fonamental, Institut de Ciències del Cosmos (ICC), Universitat de Barcelona (UB), C/ Martí i Franquès 1, 08028 Barcelona, Spain}
\affiliation[b]{Institut für Quantenphysik, Universität Ulm,\\
Albert-Einstein-Allee 11, 89081 Ulm, Germany}
\emailAdd{mfroeb@ffn.ub.edu}
\emailAdd{albert.roura@uni-ulm.de}
\emailAdd{enric.verdaguer@ub.edu}

\abstract{The Riemann correlator with appropriately raised indices characterizes in a gauge-invariant way the quantum metric fluctuations around de Sitter spacetime including loop corrections from matter fields. Specializing to conformal fields and employing a method that selects the de Sitter-invariant vacuum in the Poincar\'e patch, we obtain the exact result for the Riemann correlator through order $H^4/m_\text{p}^4$. The result is expressed in a manifestly de Sitter-invariant form in terms of maximally symmetric bitensors. Its behavior for both short and long distances (sub- and superhorizon scales) is analyzed in detail. Furthermore, by carefully taking the flat-space limit, the explicit result for the Riemann correlator for metric fluctuations around Minkowki spacetime is also obtained. Although the main focus is on free scalar fields (our calculation corresponds then to one-loop order in the matter fields), the result for general conformal field theories is also derived.}


\begin{document}
\maketitle

\section{Introduction}
\label{introduction}
De Sitter spacetime is a maximally symmetric space whose exponentially expanding part exhibits an attractor character. Any initial perturbations are redshifted away due to the exponential expansion and the vacuum left over exhibits a scale-invariant spectrum of quantum fluctuations, whose properties are closely related to the isometries of de Sitter space.
These features are central to the predictive power of cosmological inflationary models (typically characterized by a period of quasi-exponential expansion) and circumvent the need for a detailed knowledge of the initial conditions.
Indeed, as long as some mechanism gives rise to an inflating patch with a size larger than the Hubble radius, and given a minimum number of e-foldings before graceful exit from inflation, any initial excitations are effectively erased. Furthermore, the amplification of vacuum fluctuations by the quasi-exponential expansion provides a natural mechanism for the generation of primordial inhomogeneities characterized by a nearly scale-invariant power spectrum which is simply determined by the expansion rate and the deviation from exact exponential expansion. The evolution of these inhomogeneities can satisfactorily explain precise measurements of the cosmic microwave background (CMB) anisotropies and the observed large scale structure of the universe \cite{mukhanov,lyth}.

The attractor character of expanding de Sitter spacetime, which can also be particularly relevant for determining the late-time behavior of the visible universe if the current accelerated expansion is driven by a cosmological constant, is supported by so-called ``no-hair'' theorems in classical general relativity \cite{wald83,starobinsky83,friedrich86,anderson05}. On the other hand, studies at the quantum level, including the standard treatment of cosmological perturbations during inflation \cite{mukhanov92,sasaki}, have mainly been confined to a linearized treatment of both the metric perturbations and the matter fields. Nevertheless, there has recently been a growing interest in considering the quantum effects of nonlinear interactions, the corresponding corrections to the primordial spectrum and their observational implications on CMB anisotropies and large scale structure~\cite{weinberg05,sloth06,seery07}.

Significant progress has been made for interacting quantum field theories in a fixed classical de Sitter spacetime. A de Sitter-invariant state for sufficiently massive interacting theories (but without derivative interactions) was constructed in terms of its $n$-point correlation functions by analytic continuation of the Euclidean correlation functions on the hypersphere, and a quantum ``no-hair'' theorem was derived to all orders in perturbation theory~\cite{marolf10,hollands13,higuchimarolf10}. The theorem established that for any well-behaved initial state the correlation functions within a spacetime region of fixed physical size tend to those of the de Sitter-invariant vacuum at sufficiently late times. These results have been partly extended to the massless case as well~\cite{rajaraman,hollands11}. In addition, steps have been taken to include the back reaction of the quantum matter fields on the dynamics of the mean geometry within the framework of semiclassical gravity and it has been shown that the attractor character of de Sitter spacetime with respect to small metric perturbations still holds when the nonlinear back reaction from loops of arbitrary conformal matter fields is taken into account~\cite{fprv2013}.
(See also~\cite{perez-nadal08a,perez-nadal08b} for related results involving nonconformal fields and spatially isotropic metric perturbations.)

A further step in the investigation of nonlinear effects of the gravitational interaction at the quantum level is to include the quantum fluctuations of the metric perturbations around the mean geometry. These can be studied within the framework of perturbative quantum gravity regarded as an effective field theory (EFT) \cite{donoghue1994,burgess03} provided that there is a sufficiently large separation of scales between the Hubble scale $H$ and the Planck mass $m_\mathrm{p}$. Quantizing the metric perturbations is necessary in order to calculate the correlators of cosmological perturbations, and accounting for the nonlinear interaction terms amounts to computing radiative corrections beyond the standard tree-level computation.
As a matter of fact, initial calculations of one-loop corrections from matter fields found a nontrivial deviation from (nearly) scale invariance in the power spectrum of both scalar~\cite{weinberg05,chaicherdsakul07} and tensor perturbations~\cite{adshead09b} for superhorizon modes, but more detailed analysis later revealed that this was actually not the case~\cite{senatore10,frv2011a}. (Note that due to ultraviolet effects, the power spectrum defined as the equal-time limit of the spatial Fourier transform of the correlator of the metric perturbations is strictly speaking ill-defined when considering loop corrections, as shown in~\cite{frv2011a}.)
On the other hand, taking into account loops of the metric perturbations is more subtle and gives rise to infrared divergences when evaluating the correlators of cosmological perturbations at points with fixed background coordinates.
It has also been suggested for some time that radiative corrections from higher-order graviton loops%
\footnote{It should be made clear that we employ the terms ``graviton loop'' and ``graviton propagator'' to refer to the metric perturbations of scalar, vectorial and tensorial type, as usually done in this context, and not just to the tensorial ones.}
lead to a secular screening of the cosmological constant caused by the back reaction of infrared graviton modes on the dynamics of the mean-field geometry and whose effect builds up over time as more and more modes get redshifted to superhorizon scales due to the exponential expansion.
As emphasized in~\cite{gerstenlauer11,giddingssloth2011,urakawa10,tanakaurakawa2013}, whenever the metric perturbations are quantized, and particularly when considering contributions where they are involved in loops, it is important to consider observables which are not only gauge invariant under diffeomorphisms but also infrared-safe (i.e., entirely characterized by geometric properties within a region of finite physical size). 

Here we consider the connected two-point correlator of the Riemann tensor including contributions from loops of matter fields but excluding graviton loops, which can be implemented systematically within a large $N$ expansion for $N$ matter fields \cite{hu04}. The Riemann tensor with appropriately raised indices is gauge invariant for linear perturbations around de Sitter \cite{perez-nadal10} and one can see that this implies the gauge invariance of the connected Riemann correlator at order $1/N$ (i.e.\ when graviton loops are neglected). Furthermore, since the Riemann tensor characterizes the local geometry and the contributions of the metric perturbations to the physical distance between the two points at which the Riemann tensors are evaluated involve graviton loops (and are, thus, of higher order in $1/N$), the connected Riemann correlator is also an infrared-safe observable at that order.

Specifically, we compute the Riemann correlator\footnote{We always consider the connected Riemann correlator, but throughout the rest of the article we will employ the expression ``Riemann correlator'' for short.} including radiative corrections from conformal fields through order $H^4/m_p^4$, which corresponds to one-loop order for free fields and to an arbitrary number of matter loops but no internal graviton propagators for interacting conformal field theories (CFTs). An adiabatic vacuum which leads to a de Sitter invariant state for the gravitationally interacting theory and generalizes the Bunch-Davies vacuum for free fields is chosen as our asymptotic initial state, but no further assumptions about de Sitter invariance are otherwise made throughout our calculation. Remarkably, we obtain a manifestly de Sitter-invariant result for the Riemann correlator in terms of maximally symmetric bitensors. It should be emphasized that in contrast with previous studies of loop corrections to correlators of cosmological perturbations (reference~\cite{frv2011a} being an exception), our result is exact at the corresponding order and includes a detailed treatment of ultraviolet effects as well as an explicit renormalization procedure based on dimensional regularization. (See also~\cite{hawking01} for a related formal result in the Euclidean section of de Sitter space in terms of an infinite sum of products of hyperspherical tensorial harmonics.) Moreover, rather than restricting ourselves to the equal-time limit of the correlators (in a preferred background foliation), we consider arbitrary pairs of spacetime points, which is essential in order to show the de Sitter invariance of our results.

Although we mainly focus on the case of free massless scalar fields with conformal curvature coupling, our results can be easily extended to any CFT (even interacting ones) and the general expression for the Riemann correlator is provided in section~\ref{general_CFT}. This is possible because the result of integrating out the matter fields is given by a nonlocal effective action whose form is the same for any CFT except for the values of two constant coefficients, as explained in~\cite{fprv2013} and briefly reviewed in section \ref{2pf_action} below. Integrating out the metric perturbations and obtaining the loop-corrected Riemann correlator is then completely analogous to the case of the free scalar fields.
As a separate by-product, we also compute the Riemann correlator in Minkowski spacetime by carefully taking the flat-space limit of our result in de Sitter.
To the best of our knowledge, this result for the Riemann correlator in Minkowski spacetime including loop corrections from conformal fields has not been obtained previously either. There are results in the literature for the metric correlator \cite{tomboulis77,martin00} and the correlator of the Einstein tensor \cite{martin00}, which is closely related to the stress tensor correlator \cite{osbornshore}, but not for the full Riemann correlator.

The article is organized as follows. In section~\ref{metricpert} we review the decomposition of the metric perturbations, taking into account the symmetries of the background spatial sections, into scalar, vectorial and tensorial perturbations. We write the perturbations in terms of spatial Fourier modes and fix the gauge for quantization. In section~\ref{2pf} we first briefly review the closed-time-path (CTP) effective action method for computing, in the Poincaré patch, the correlators of the different types of metric perturbations with respect to the de Sitter-invariant vacuum of the interacting theory, selected by the $\mathi \epsilon$ prescription that we employ. For the scalar and vectorial modes, which are constrained degrees of freedom, the calculation is performed from the start and we compute the correlators up to order $H^4/m_p^4$ (which corresponds to one-loop order for free matter fields). For the tensorial modes the calculation is much more involved, but we can take advantage of the result previously obtained in~\cite{frv2011a}. The result for the metric correlator is of course gauge dependent and does not exhibit de Sitter invariance explicitly, but it is invariant under spatial rotations and translations as well as under a simultaneous rescaling of the spatial and conformal time coordinates, which are necessary conditions for de Sitter invariance. In section \ref{bitensors} we review the formalism of de Sitter-invariant bitensors developed by Allen and Jacobson \cite{allenjacobson86} and we adapt it to the spatially flat coordinates of the Poincaré patch in de Sitter space. Section \ref{curvature} contains the main results of the paper: we compute the connected two-point correlator of the Riemann tensor $R^{ab}{}_{cd}$ by taking suitable derivatives of the metric correlator computed in section \ref{2pf}. We are able to write the result in a manifestly de Sitter-invariant form in terms of covariant derivatives of the invariant interval $Z(\mu)$, which is a function of the geodesic distance $\mu(x,x')$ between the two points. We also provide an alternative expression of the correlator  in terms of the de Sitter-invariant bitensors introduced by Allen and Jacobson. Moreover, given the decomposition of the Riemann tensor in terms of Weyl and Ricci tensors, we write the explicit results for the Weyl and Ricci correlators as well as the Weyl-Ricci correlator and study both their short and long-distance (superhorizon) behavior. These results are extended to general CFTs in section~\ref{general_CFT} and the flat-space limit of all the correlators is provided in section~\ref{curvature_flat}. In section \ref{comparison} we compare with previous results for the Weyl correlator at tree level \cite{kouris,morawoodard12a,morawoodard12b} and the Ricci correlator at one-loop order, which can be directly related to the stress tensor correlator of the matter fields \cite{osbornshore,perez-nadal10}. Finally, we discuss our main results and perspectives for future work in section~\ref{discussion}.
Some technical details and long expressions are included in the Appendices.

We use the ``+++'' sign convention of~\cite{mtw}. According to our notation Latin indices range over space and time, while Greek indices are used for spatial components. Furthermore, we work in natural units with $c = \hbar = 1$ and set $\kappa^2 = 16 \mathpi G_\text{N} = 16 \mathpi / m_p^2$\,, where $G_\text{N}$ is Newton's gravitational constant. Covariant derivatives with respect to the background de Sitter metric are denoted by $\hat{\nabla}_a$ or a semicolon. In general, objects referring to the de Sitter background are indicated with a hat over the symbol, while objects which include perturbations around this metric are marked with a tilde. The spatial Fourier transform of a given quantity is in turn denoted by an overbar.

\section{Decomposition of the metric perturbations}
\label{metricpert}

In this section we decompose the metric perturbations into different sectors according to their behavior under spatial transformations as was first done in \cite{lifshitz,bardeen}, and fix the gauge. In contrast to the Fadeev-Popov gauge-fixing procedure, which is implemented at the level of the generating functional and leads to the introduction of additional ghost fields to cancel residual unphysical modes, we fix the gauge completely proceeding in a way analogous to that advocated in \cite{maldacena} and which is essentially equivalent to working in reduced phase-space. This ensures that only physical modes contribute. In addition, the approach is not affected by recent controversy about the validity of (certain) average gauges in de Sitter spacetime \cite{woodard2009,moratsamiswoodard2012,higuchi2011,faizal2012,higuchi2012,morrison2013}.

In the Poincaré patch of de Sitter spacetime, the de Sitter metric with spatially flat sections can be written using Cartesian coordinates as $\hat{g}_{ab} = a^2(\eta) \eta_{mn}$, where $\eta_{mn}$ is the Minkowski metric, $-\infty < \eta \leq 0$ is the conformal time and $a(\eta) = (-H\eta)^{-1}$. To this metric we add perturbations $h_{mn}$. The resulting metric reads
\begin{equation}
\label{metric_def_pert}
\tilde{g}_{mn} = a^2(\eta) g_{mn} = a^2(\eta) \left( \eta_{mn} + \kappa h_{mn} \right) = \hat{g}_{mn} + a^2(\eta) \kappa h_{mn} \eqend{.}
\end{equation}
The metric perturbation $h_{mn}$ is sometimes called the ``pseudo-graviton field'' \cite{pseudograviton}. In general we will use the tilde over a symbol to refer to terms defined with the perturbed physical metric $\tilde{g}_{ab}$, while hats over a symbol denote quantities defined in the de Sitter background $\hat{g}_{ab}$. Under an infinitesimal coordinate transformation of the full metric generated by a vector field $\psi_m$, the physical perturbation $a^2 h_{mn}$ changes by the gauge transformation
\begin{equation}
a^2 h_{mn} \to a^2 h_{mn} + \hat{\nabla}_m \hat{\xi}_n + \hat{\nabla}_n \hat{\xi}_m \eqend{,}
\end{equation}
which upon rescaling of the vector field $\hat{\xi}_m = a^2 \xi_m$ gives
\begin{equation}
\label{gauge_trafo}
h_{mn} \to h_{mn} + \partial_m \xi_n + \partial_n \xi_m + \frac{2}{\eta} \eta_{mn} \xi_0 \eqend{.}
\end{equation}

To fix the gauge, we first decompose the perturbation $h_{mn}$ according to its transformation properties under spatial diffeomorphisms. The spatial part $h_{\alpha\beta}$ ($\alpha,\beta = 1,2,3$) can be decomposed as
\begin{equation}
\label{perturbation_spatial_decompose}
h_{\alpha\beta} = h^\text{TT}_{\alpha\beta} + 2 \partial_{(\alpha} w^\text{T}_{\beta)} + \partial_\alpha \partial_\beta \sigma + \tau \delta_{\alpha\beta} \eqend{,}
\end{equation}
where $\delta^{\mu\alpha} \partial_\mu h^\text{TT}_{\alpha\beta} = 0 = \delta^{\alpha\beta} h^\text{TT}_{\alpha\beta}$ and $\delta^{\alpha\beta} \partial_\alpha w^\text{T}_\beta = 0$. The temporal components similarly  can be decomposed as
\begin{equation}
\label{perturbation_temporal_decompose}
h_{0\alpha} = v^\text{T}_\alpha + \partial_\alpha \psi \eqend{,} \qquad h_{00} = \varphi \eqend{,}
\end{equation}
where $\delta^{\alpha\beta} \partial_\alpha v^\text{T}_\beta = 0$. In total, we have four scalars $\varphi$, $\psi$, $\sigma$, $\tau$, two transverse vectors $w_\alpha$ and $v_\alpha$ with two kinematical degrees of freedom each, and a transverse traceless tensor $h^\text{TT}_{\alpha\beta}$ with also two kinematical degrees of freedom, which amount to 10 kinematical degrees of freedom as expected for a symmetric $4\times4$ matrix.

Decomposing also the vector field $\xi_m$ as
\begin{equation}
\xi_\alpha = \chi^\text{T}_\alpha + \partial_\alpha \xi \eqend{,}
\end{equation}
where $\delta^{\alpha\beta} \partial_\alpha \chi^\text{T}_\beta = 0$, we can see that the different components change under a gauge transformation as follows:
\begin{equation}
\begin{aligned}
h^\text{TT}_{\alpha\beta} &\to h^\text{TT}_{\alpha\beta} & w^\text{T}_\alpha &\to w^\text{T}_\alpha + \chi^\text{T}_\alpha \\
\sigma &\to \sigma + 2 \xi & \tau &\to \tau + \frac{2}{\eta} \xi_0 \\
v^\text{T}_\alpha &\to v^\text{T}_\alpha + \chi^{\text{T}\, \prime}_\alpha & \psi &\to \psi + \xi' + \xi_0 \\
\varphi &\to \varphi + 2 \xi'_0 - \frac{2}{\eta} \xi_0 \eqend{.} & &
\end{aligned}
\end{equation}
Choosing $\chi^\text{T}_\alpha$, $\xi$ and $\xi_0$ appropriately, we can therefore set $w_\alpha$, $\sigma$ and $\tau$ to zero. Then, the spatial part -- now in the transverse traceless gauge -- describes the dynamical, physical perturbations, which are gravitational waves with two independent polarizations.

We take now the Fourier transform with respect to the spatial arguments
\begin{equation}
h_{mn}(\eta,\vec{x}) = \int \bar{h}_{mn}(\eta, \vec{p}) \mathe^{\mathi \vec{p} \vec{x}} \frac{\total^3 p}{(2\mathpi)^3} \eqend{.}
\end{equation}
Here we introduced the notation that will be used throughout the article of using an overbar over a quantity to indicate the spatial Fourier transform of that object. After fixing the gauge $w^\text{T}_\alpha = 0 = \sigma = \tau$, the metric perturbation \eqref{perturbation_spatial_decompose}-\eqref{perturbation_temporal_decompose} reads in terms of its Fourier modes as
\begin{equation}
\label{metricpert_fixed_decomp}
\bar{h}_{mn} = \delta^0_m \delta^0_n \bar{\varphi} + 2 \delta^0_{(m} \left( \bar{v}^\text{T}_{n)} + \mathi p_{n)} \bar{\psi} \right) + \bar{h}^\text{TT}_{mn} \eqend{,}
\end{equation}
where now
\begin{equation}
\bar{v}_0 = p_0 = p^m \bar{v}^\text{T}_m = p^m \bar{h}^\text{TT}_{mn} = \bar{h}^\text{TT}_{0m} = \eta^{mn} \bar{h}^\text{TT}_{mn} = 0 \eqend{.}
\end{equation}
In the next section, we will quantize the metric perturbations. As we will see, whereas the tensorial perturbations are true dynamical degrees of freedom of the gravitational field, the scalar and vectorial parts are constrained degrees of freedom since they are determined by the matter content of the theory, i.e., by the stress tensor through the Einstein equations. Thus, at the operator level these perturbations could be eliminated in favor of the matter fields, as advocated in cosmology by Maldacena \cite{maldacena}. However, in our case it is more convenient to deal with these constrained perturbations in a different way, which is nevertheless fully equivalent to their elimination at the operator level.

\section{Two-point functions of the metric perturbations}
\label{2pf}

We will consider the quantization of the metric perturbations $h_{ab}$ around a spatially flat Friedmann-Lemaître-Robertson-Walker (FLRW) background within the framework of perturbative quantum gravity regarded as an effective field theory \cite{donoghue97,burgess03}, and will calculate the two-point correlation function of the metric perturbations including the radiative corrections from $N$ massless, conformally coupled scalar fields. The fields have a vanishing expectation value and the evolution of the background spacetime is driven by a cosmological constant. Furthermore, the correlation function will be computed to leading order in $1/N$, which excludes any contributions from graviton loops since those are of higher order in $1/N$. In that case one only needs counterterms in the gravitational action up to quadratic order in the curvature (otherwise higher curvature terms as well as higher-order terms involving the matter fields would be required) and the total bare action is, therefore, given by
\begin{equation}
\label{EHaction}
\begin{split}
S[\tilde{g}, \tilde{\phi}] &= \frac{N}{\bar{\kappa}^2} \int \left( \tilde{R} - 2 \Lambda \right) \sqrt{-\tilde{g}} \total^n x + N a_1 \int \left( \tilde{R}^{abcd} \tilde{R}_{abcd} - \tilde{R}^{ab} \tilde{R}_{ab} \right) \sqrt{-\tilde{g}} \total^n x \\
&\quad+ N a_2 \int \tilde{R}^2 \sqrt{-\tilde{g}} \total^n x - \frac{1}{2} \sum_{k=1}^N \int \left[ \left( \tilde{\nabla}^m \tilde{\phi}_k \right) \left( \tilde{\nabla}_m \tilde{\phi}_k \right) + \xi(n) \tilde{R} \tilde{\phi}_k^2 \right] \sqrt{-\tilde{g}} \total^n x \eqend{.}
\end{split}
\end{equation}
All the parameters in the purely gravitational part of the action ($\bar\kappa$, $a_1$, $a_2$ and $\Lambda$) are in principle bare parameters. On the other hand, working at leading order in $1/N$ the action for the matter fields is finite and the parameter $\xi(n) = (n-2)/[4(n-1)]$ has been chosen so that the matter sector is conformally invariant in arbitrary dimensions at the classical level. We consider the action in $n$ dimensions so that dimensional regularization can be employed.
Moreover, we have introduced a rescaled gravitational coupling $\bar{\kappa}^2 = N \kappa^2$ which enables a useful organization of the Feynman diagrams in terms of a large $N$ expansion for $N$ matter fields. Graviton loops are then suppressed in comparison to matter loops by additional powers of $1/N$ and we can neglect them when calculating the correlation functions to leading order in $1/N$ \cite{hu04}.
Finally, in order to take advantage of the conformal flatness of the background spacetime in our calculation, it is convenient to rescale the scalar fields as $\tilde{\phi} = a^{1-\frac{1}{2}n} \phi$ when carrying out explicit computations. Sometimes it is also convenient to rewrite the first counterterm quadratic in the curvature in terms of the square of the Weyl tensor and the integrand of the four-dimensional Euler invariant, $\tilde{\mathcal{E}}_4 = \tilde{R}^{abcd} \tilde{R}_{abcd} - 4 \tilde{R}^{ab} \tilde{R}_{ab} + \tilde{R}^2$, as follows:
\begin{equation}
\label{counterterm}
\begin{split}
&\int \left( \tilde{R}^{abcd} \tilde{R}_{abcd} - \tilde{R}^{ab} \tilde{R}_{ab} \right) \sqrt{-\tilde{g}} \total^n x = - \left( \frac{1}{2} - \frac{3}{4} (n-4) \right) \int \tilde{\mathcal{E}}_4 \sqrt{-\tilde{g}} \total^n x \\
&\qquad+ \frac{n-4}{12} \int \!\! \tilde{R}^2 \sqrt{-\tilde{g}} \total^n x + \left( \frac{3}{2} - \frac{3(n-4)}{4} \right) \int \tilde{C}^{abcd} \tilde{C}_{abcd} \sqrt{-\tilde{g}} \total^n x + \bigo{(n-4)^2}
\eqend{,}
\end{split}
\end{equation}
where the expression for the Weyl tensor in $n$ dimensions is to be considered here.

Since we want to calculate true expectation values rather than transition matrix elements, we need to use the ``in-in'' (or CTP) formalism. This is a generalization of the standard ``in-out'' formalism where one introduces two copies of each field (distinguished by the superscript~$\pm$) or, alternatively, one considers an integration contour for the time integrals that goes from the (asymptotic) initial time to some sufficiently late time and back to the initial time \cite{cooper94}. We give here a short overview of the method employed (a more detailed description is provided in \cite{frv2011a}) and for notational simplicity we suppress the tensor indices. One starts by introducing the CTP generating functional $Z[J^\pm]$, which depends on two sources $J^\pm$ and is given by
\begin{equation}
\label{ctp_generating_functional_as_path_integral}
\begin{split}
Z[J^\pm] &= \tr\left[ \left( \mathcal{T} \mathe^{\mathi \int J^+(x) \op{h}(x) \total^n x} \right) \op{\rho} \left( \widetilde{\mathcal{T}} \mathe^{-\mathi \int J^-(x) \op{h}(x) \total^n x} \right) \right] \\
&= \int \delta\big[ \phi^+(T) - \phi^-(T) \big] \delta\big[ h^+(T) - h^-(T) \big] \delta\big[ h^\pm(t_0) - h^\pm_0 \big] \delta\big[ \phi^\pm(t_0) - \phi^\pm_0 \big] \times \\
&\quad\times \bra{\phi^+_0, h^+_0} \op{\rho} \ket{\phi^-_0, h^-_0} \mathe^{\mathi ( S[h^+, \phi^+] + \int \! J^+ h^+ \total^n x ) - \mathi ( S[h^-, \phi^-] + \int \! J^- h^- \total^n x )} \mathcal{D} \phi^\pm \mathcal{D} h^\pm \mathcal{D} \phi_0^\pm \mathcal{D} h_0^\pm \eqend{,}
\end{split}
\end{equation}
where $\mathcal{T}$ and $\widetilde{\mathcal{T}}$ denote respectively time ordering and anti-time ordering, the density matrix $\op{\rho}$ describes the state of the system at the initial time $t_0$, and $\phi_0^\pm$ and $h_0^\pm$ are field configurations at the initial time $t_0$ which are also integrated over. This time can be finite, although one would then need to consider an appropriately dressed state for the interacting theory. The final time $T$ in the generating functional is arbitrary as long as it is larger than all the times of interest, i.e., the times appearing in the correlation function that we want to calculate. In practice, a convenient procedure is to assume a factorized initial density matrix $\bra{\phi^+_0, h^+_0} \op{\rho} \ket{\phi^-_0, h^-_0} = \bra{\phi^+_0} \op{\rho}_\phi \ket{\phi^-_0} \bra{h^+_0} \op{\rho}_h \ket{h^-_0}$ and perform the functional integral over the matter fields first. The selection of the right asymptotic initial state for the interacting theory, which includes correlations between the matter fields and the metric perturbations, can still be implemented by making use of a suitable $\mathi\epsilon$ prescription in the subsequent time integration contours and taking $t_0 \to -\infty$ \cite{frv2011a}.

The two-point functions, which are true expectation values with respect to the ``in'' state, are then obtained by functionally differentiating the generating functional $Z[J^\pm]$ with respect to the sources:
\begin{equation}
\label{propagator_functional_derivative}
\bra{\text{in}} \mathcal{P} h_A(x) h_B(x') \ket{\text{in}} = \left[ Z^{-1}[J^\pm] \frac{\delta^2 Z[J^\pm]}{\mathi \delta J_A(x) \, \mathi \delta J_B(x')} \right]_{J = 0} = \mathi G_{AB}(x, x') \eqend{,}
\end{equation}
where the indices $A$ and $B$ can take the values $\pm$. The path ordering denoted by $\mathcal{P}$ is a generalization of the time ordering $\mathcal{T}$ and the anti-time ordering $\tilde{\mathcal{T}}$. Particularizing the four possible values of the index pair $AB$, we have
\begin{equation}
\label{ctp_propagator}
\begin{split}
G_{++}(x, x') &= - \mathi \bra{\text{in}} \mathcal{T} h(x) h(x') \ket{\text{in}} \eqend{,} \qquad G_{-+}(x, x') = - \mathi \bra{\text{in}} h(x) h(x') \ket{\text{in}} \eqend{,} \\
G_{--}(x, x') &= - \mathi \bra{\text{in}} \tilde{\mathcal{T}} h(x) h(x') \ket{\text{in}} \eqend{,} \qquad G_{+-}(x, x') = - \mathi \bra{\text{in}} h(x') h(x) \ket{\text{in}} \eqend{.}
\end{split}
\end{equation}

When performing the functional integrals for the matter fields in equation~\eqref{ctp_generating_functional_as_path_integral}, the integrals over the scalar field $\phi$ are convergent in $n \neq 4$ dimensions, but exhibit ultraviolet divergences as $n \to 4$, which are canceled by the counterterms in the bare action~\eqref{EHaction} (for massless fields only counterterms quadratic in the curvature tensors are necessary). One is then left with
\begin{equation}
\label{ctp_generating_functional_only_h}
\begin{split}
Z[J^\pm] &= \int \delta\big[ h^+(T) - h^-(T) \big] \delta\big[ h^\pm(t_0) - h^\pm_0 \big] \times \\
&\qquad\times \bra{h^+_0} \op{\rho}_h \ket{h^-_0} \mathe^{\mathi ( S_\text{G}[h^+] + \int \! J^+ h^+ \total^n x ) - \mathi ( S_\text{G}[h^-] + \int \! J^- h^- \total^n x ) + \mathi \Sigma[h^\pm]} \mathcal{D} h^\pm \mathcal{D} h_0^\pm \eqend{,}
\end{split}
\end{equation}
where $S_\text{G}[h]$ corresponds to the gravitational part of the original action~\eqref{EHaction} and $\Sigma[h^\pm]$ results from the functional integration of the matter fields. Both $S_\text{G}[h]$ and $\Sigma[h^\pm]$ are already renormalized quantities, after the cancellation of the divergences arising in the functional integration with the counterterms in the bare gravitational action has taken place.
They can be combined into the following quantity, which is often called the CTP effective action \cite{hu08}:
\begin{equation}
\label{seff}
S_\text{eff}[h^\pm] = S_\text{G}[h^+] - S_\text{G}[h^-] + \Sigma[h^\pm]
= S_0[h^+] - S_0[h^-] + S_\text{int}[h^\pm] \eqend{,}
\end{equation}
where we have separated in the second equality the free part $S_0[h^\pm]$ of the gravitational action (quadratic in $h^\pm$ and of order $\kappa^0$) and all the remaining terms, including the nonlocal terms in $\Sigma[h^\pm]$, which are denoted by $S_\text{int}[h^\pm]$.
The generating functional \eqref{ctp_generating_functional_only_h} can then be rewritten as
\begin{equation}
\label{ctp_generating_functional_int}
Z[J^\pm] = \mathe^{\mathi S_\text{int}[\mp \mathi \delta/\delta J^\pm]}\, Z_0[J^\pm]
\eqend{,}
\end{equation}
where $Z_0[J^\pm]$ is the generating functional associated with the free part. This is a useful way of organizing the calculation which allows (by expanding the exponential) a simple derivation of the Feynman rules for general interacting theories and is directly applicable to both the ``in-out'' and CTP formalisms.
When computing connected correlation functions to leading-order in $1/N$, we only need to consider terms in $S_\text{eff}[h^\pm]$ which are linear and quadratic in $h$ since terms with higher powers of $h$ give rise to contributions of higher order in $1/N$ \cite{hu04}. (There is also an irrelevant term which does not depend on the perturbation $h$.) If we demand that the background metric $a^2(\eta) \eta_{ab}$ is a solution of the semiclassical Einstein equation\footnote{\label{Lambda_eff}A small shift of the background cosmological constant $\Lambda$ is thus introduced, leading to a new $\Lambda_\text{eff}$ which defines a quantum corrected Hubble parameter $H = \sqrt{\Lambda_\text{eff}/3}$ in $a(\eta)$; see~\cite{frv2011a} for details.}, the linear terms cancel out and we are left with a purely quadratic expression.
At this order the the renormalized CTP effective action becomes
\begin{equation}
\label{seff_powers_kappa}
S_\text{eff}[h^\pm] = S_0[h^+] - S_0[h^-] + \kappa^2 S_2[h^\pm]  \eqend{,}
\end{equation}
where only terms quadratic in $h$ should be included and the result has been expanded in powers of $\kappa$.

\begin{figure}[ht]
\begin{center}
\includegraphics[width=\textwidth]{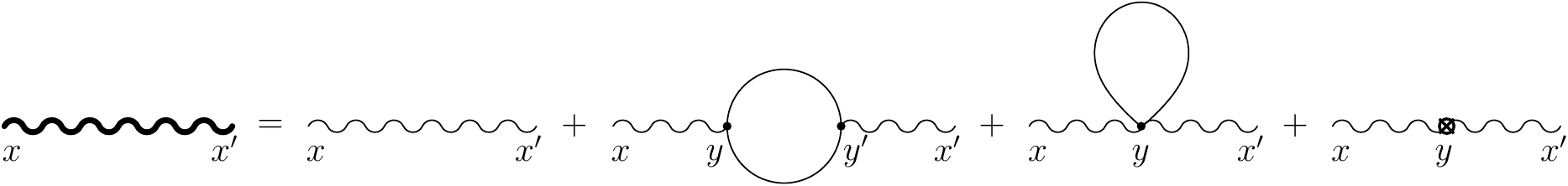}
\end{center}
\caption{Feynman-Stueckelberg diagram corresponding to our calculation of the two-point quantum correlator of the metric perturbations up to order $\kappa^2$ for a massless, conformally coupled free scalar field. Wiggly lines correspond to metric perturbations and solid lines to scalar fields.}
\label{fig:free}
\end{figure}

From equation~\eqref{seff_powers_kappa} we have $S_\text{int}[h^\pm] = \kappa^2 S_2[h^\pm]$ and one can then obtain the correlation functions up to order $\kappa^2$ (and leading order in $1/N$) by expanding the exponential in equation \eqref{ctp_generating_functional_int} up to that order (which corresponds, for free fields, to one-loop order in the matter fields):
\begin{equation}
\label{ctp_generating_functional_trunc}
Z[J^\pm] = \left( 1 + \kappa^2 S_2\left[\mp \mathi\, \delta/\delta J^\pm \right] + \bigo{\kappa^4} \right) Z_0[J^\pm]
\eqend{.}
\end{equation}
The generating functional for the free part is given by
\begin{equation}
\label{ctp_generating_functional_free}
Z_0[J^\pm] = \exp \left[-\frac{\mathi}{2} \int \int J^A(x) G^0_{AB}(x,x') J^B(x') \total^4 x \total^4 x' \right] \eqend{,}
\end{equation}
where $G^0_{AB}$ corresponds to the free part of the two-point correlation functions \eqref{propagator_functional_derivative}. Substituting equation \eqref{ctp_generating_functional_free} into equation \eqref{ctp_generating_functional_trunc} and functionally differentiating with respect to the sources $J^\pm$, we obtain the following result for the two-point functions~\eqref{propagator_functional_derivative} up to order $\kappa^2$:
\begin{equation}
\label{propagator_first_order_integral}
G_{AB}(x, x') = G^0_{AB}(x,x') - \kappa^2 \iint G^0_{AM}(x, y) V_{MN}(y, y') G^0_{NB}(y', x') \total^4 y' \total^4 y + \bigo{\kappa^4} \eqend{,} \quad
\end{equation}
where $V$ is the amputated one-loop contribution from $S_2$, given by
\begin{equation}
\label{perturbation_v_definition}
V = \begin{pmatrix} \frac{\delta^2}{\delta h^+ \delta h^+} & \frac{\delta^2}{\delta h^+ \delta h^-} \\ \frac{\delta^2}{\delta h^- \delta h^+} & \frac{\delta^2}{\delta h^- \delta h^-} \end{pmatrix} S_2[h^+, h^-] \eqend{.}
\end{equation}
The Feynman-Stueckelberg diagrams corresponding to equation~\eqref{propagator_first_order_integral} are shown in figure~\ref{fig:free}.

Also known as the CTP propagator of the free theory, $G^0_{AB}$ satisfies the equation
\begin{equation}
\label{propagator_definition}
\int A^0_{AM}(x, y) G^0_{MB}(y, x') \total^4 y = \delta^4(x-x') \delta_{AB} \eqend{,}
\end{equation}
with the tree-level kinetic operator $A^0_{AM}$ defined as
\begin{equation}
\label{a0_definition}
A^0_{MN} = \begin{pmatrix} \frac{\delta^2 S_0[h^+]}{\delta h^+ \delta h^+} & 0 \\ 0 & - \frac{\delta^2 S_0[h^-]}{\delta h^- \delta h^-} \end{pmatrix} \eqend{.}
\end{equation}
Note, however, that equation \eqref{propagator_definition} does not specify completely the free propagator $G^0_{AB}$ and one needs to provide, in addition, appropriate boundary conditions which determine the operator ordering of the different components as well as the quantum state of the field. 
In equations \eqref{ctp_generating_functional_free}-\eqref{a0_definition}, capital Latin indices take [as in equation~\eqref{propagator_functional_derivative}] the values $\pm$, and repeated indices are summed over.

Equation~\eqref{propagator_first_order_integral} involves time integrals between the initial time $t_0$ and the final time $T$ with a suitable $\mathi \epsilon$ prescription that selects the true asymptotic vacuum state of the interacting theory. As mentioned above, this consists in sending the initial time $t_0$ to $-\infty$ along a slightly imaginary direction:
\begin{equation}
t_0^\pm \to -\infty(1 \mp \mathi\epsilon) \qquad\text{with }\epsilon > 0 \eqend{.}
\end{equation}
Whether we have to take $t_0^+$ or $t_0^-$ depends on the corresponding index ($M$ or $N$) in equation~\eqref{propagator_first_order_integral}, e.g., if $M = +$, the integral over $y$ runs from $t_0^+$ to $T$.

\subsection{The CTP effective action}
\label{2pf_action}

The renormalized effective action $S_\text{eff}[h^\pm]$ for the free conformal field model specified by equation \eqref{EHaction} has been calculated up to quadratic order in $h^\pm$  in \cite{camposverdaguer94,camposverdaguer96}.
In fact, as emphasized in section~8 of~\cite{fprv2013}, the result is essentially the same for any CFT and is given at that order by
\begin{equation}
\label{seff_cft}
S_\text{eff}[h^\pm] = S_\text{G}[h^+] - S_\text{G}[h^-] + \kappa^2 S^\text{K}_2 [h^\pm] \eqend{,}
\end{equation}
where $S_\text{G}$ takes the following simple form \cite{fprv2013,mazur01}:
\begin{equation}
\label{sG_cft}
\begin{split}
S_\text{G}[h] &= \frac{1}{\kappa^2} \int \left( \tilde{R} - 2 \Lambda \right) \sqrt{-\tilde{g}} \total^4 x + \beta \int \tilde{R}^2 \sqrt{-\tilde{g}} \total^4 x + b \int \ln a\, C^{abcd} C_{abcd} \sqrt{-g} \total^4 x \\
&\quad+ \lim_{n \to 4}\, \frac{b'}{n-4} \left( \int \tilde{\mathcal{E}}_4 \sqrt{-\tilde{g}} \total^n x
- \int \mathcal{E}_4 \sqrt{-g} \total^n x \right) \eqend{.}
\end{split}
\end{equation}
Note that the difference between the integrals of the Euler densities for two conformally related spacetimes is of order $(n-4)$ and the contribution proportional to $b'$ in equation \eqref{sG_cft} is, thus, always finite. Furthermore, since $\mathcal{E}_4$ vanishes through quadratic order for metric perturbations around Minkowski spacetime, only the integral of $\tilde{\mathcal{E}}_4$ contributes up to that order in our case.
On the other hand, the last term on the right-hand side of equation~\eqref{seff_cft}, $S^\text{K}_2 [h^\pm]$, is a nonlocal term quadratic in the linearized Weyl tensor
which coincides with $\Sigma[h^\pm]$ at quadratic order and is given by
\begin{equation}
\label{effective_action_s2k}
\kappa^2 S_2^\text{K}[h^\pm] = 
b \iint C_M^{abcd}(x) \begin{pmatrix} K^+(x-y; \bar{\mu}) & K(x-y) \\ K(y-x) & -K^-(x-y; \bar{\mu}) \end{pmatrix}_{MN}\! C_{Nabcd}(y) \total^4 x \total^4 y \eqend{,}
\end{equation}
where $C_{abcd}(x)$ is the linearized Weyl tensor evaluated for the perturbation $\kappa\, h_{ab}$ around the flat-space metric $\eta_{ab}$ [i.e., without the conformal factor $a(\eta)$] and the kernels $K$ are given in terms of the following spacetime Fourier transforms:
\begin{equation}
\label{kernels_def}
\begin{split}
K(x) &= - \mathi \mathpi \int \Theta(-p^2) \Theta(-p^0) \mathe^{\mathi p x} \frac{\total^4 p}{(2\mathpi)^4} \\
K^\pm(x; \mu) &= \frac{1}{2} \int \left[ - \ln \abs{\frac{p^2}{\mu^2}} \pm \mathi \mathpi \Theta(-p^2) \right] \mathe^{\mathi p x} \frac{\total^4 p}{(2\mathpi)^4} \eqend{,}
\end{split}
\end{equation}
with $\mu$ being some renormalization scale.
In \cite{camposverdaguer94,camposverdaguer96} a value of the renormalization scale $\bar{\mu}$ was chosen so that the renormalized coupling constant $a_1^\text{ren}(\bar{\mu})$ in equation~\eqref{EHaction} vanishes. To shorten the intermediate expressions, we will also make this choice. However, once we obtain the final result in section~\ref{curvature}, we will specify the small changes needed so that the result is valid for an arbitrary $\mu$. More specifically, from equations~\eqref{EHaction}--\eqref{counterterm} together with \eqref{effective_action_s2k}--\eqref{kernels_def} we have the following relation between two different values of the renormalization scale $\mu_0$ and $\mu$:
\begin{equation}
N a_1^\text{ren}(\mu) = N a_1^\text{ren}(\mu_0) - 
\frac{2}{3}\,b\, \ln \left( \frac{\mu}{\mu_0} \right) \eqend{,}
\end{equation}
so that for an arbitrary value $\mu$ of the renormalization scale we have to make the replacement $b \ln \bar\mu \to (3/2) N a_1^\text{ren}(\mu) + b \ln \mu$. This guarantees that the effective action $S_\text{eff}[h^\pm]$ is invariant under the action of the renormalization group.
\begin{figure}[ht]
\begin{center}
\includegraphics[width=0.85\textwidth]{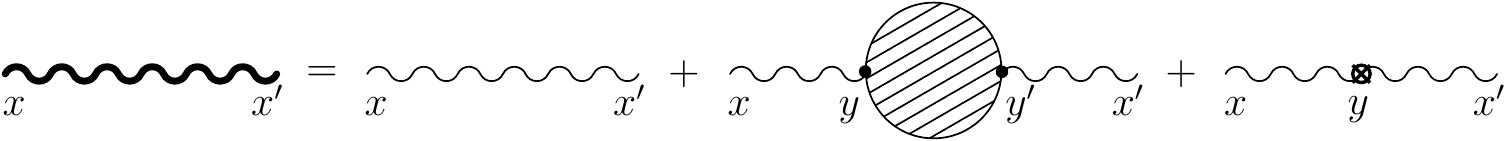}
\end{center}
\caption{Diagram corresponding to our calculation of the two-point quantum correlator of the metric perturbations up to order $\kappa^2$ for a general CFT. The filled circle represents contributions with an arbitrary number of matter loops but no internal graviton propagators, and the last diagram is the contribution from the counterterms, which after renormalization give rise to finite contributions [multiplied by $\beta$, $b$ and $b'$ in equation~\eqref{sG_cft}].}
\label{fig:cft}
\end{figure}

Different CFTs will correspond to different values of the constants $b$ and $b'$ [for instance, $b = 3 \alpha / 2$ and $b' = - \alpha / 2$ with $\alpha = N/(2880\pi^2)$ for the free scalar field model specified in equation \eqref{EHaction}] and $\beta$ is a free parameter, but the expressions for $S_\text{eff}$ are otherwise the same. This means that the results for the Riemann correlator through order $\kappa^4$ obtained in section~\ref{curvature} will also be valid for general CFTs, even interacting ones. For interacting fields this will include radiative corrections with an arbitrary number of matter loops but no graviton loops (which would give rise to contributions of higher order in $1/N$) nor any internal graviton propagators connecting different matter loops (which correspond to contributions of higher order in $\kappa^2$). The corresponding Feynman-Stueckelberg diagram is shown in figure~\ref{fig:cft}.

The effective action for conformal fields~\eqref{seff_cft} has been previously computed in a number of references \cite{starobinsky81,camposverdaguer96,hawking01}, sometimes neglecting the conformally invariant nonlocal term \cite{riegert84,fradkin84,mazur01,fabris01}. This has been employed to derive the linearized semiclassical Einstein equation (obtained by functionally differentiating the effective action with respect to $h_{ab}$) and study the evolution of the mean spacetime geometry when considering linear perturbations around a conformally flat background metric~\cite{starobinsky81,horowitz80,horowitz82,camposverdaguer96,anderson09,fabris12,fprv2013}.
(See also~\cite{martin00,gorbar03} for related results involving massive (scalar) fields coupled to metric perturbations around a Minkowski background.)
It should be emphasized that the effective action (and the linearized semiclassical equation) essentially corresponds to the amputated version of the diagrams appearing in  figures~\ref{fig:free} and \ref{fig:cft}, i.e., without the external legs involving the free propagators of the metric perturbations.
Instead, the main objects of study in the present paper are the correlator for the quantum fluctuations of the metric perturbations around the mean background geometry, with associated Feynman-Stueckelberg diagrams depicted in figures~\ref{fig:free} and \ref{fig:cft}, as well as the corresponding Riemann correlator.
In contrast to the situation for a Minkowski background, the exact computation of the two vertex integrals needed in order to obtain the diagrams with external legs from the amputated diagrams is highly nontrivial in de Sitter due to the time-dependence of the background metric and the more complex expression for the graviton propagator. This also makes the computation of the Riemann correlator (an eighth rank bitensor) and the presentation of the result in a manifestly de Sitter-invariant form, carried out in section~\ref{curvature}, rather involved.

\subsubsection{Decomposition into scalar, vectorial and tensorial parts}
\label{seff_decomposition}

From this point on we specialize to the case of free scalar fields by taking the particular values specified above for the parameters $b$ and $b'$, and will return to the case of general CFTs in section~\ref{general_CFT}. In the rest of this subsection we decompose the effective action into scalar, vectorial and tensorial parts. This will be done for the local part $S_0 + S^\text{G}_2$ first, and then for the nonlocal part $S^\text{K}_2$.

Inserting the decomposition of the metric perturbation \eqref{metricpert_fixed_decomp}, we see that the tensorial, vectorial and scalar perturbations decouple. Specifically, when expanded in powers of $\kappa$ as done in equation~\eqref{seff_powers_kappa}, the effective action \eqref{seff_cft}, which only contains terms quadratic in $h^\pm$, can be written as
\begin{equation}
\label{effective_action_decomp}
\begin{split}
S_0[h] &= S_0^\text{S}[\varphi,\psi] + S_0^\text{V}[v^\text{T}_m] + S_0^\text{TT}[h^\text{TT}_{mn}] \eqend{,} \\
S_2[h^\pm] &= S^\text{G}_2[h^+] - S^\text{G}_2[h^-] + S^\text{K}_2[h^\pm] \eqend{,} \\
S^\text{G}_2[h] &= S^\text{G,S}_2[\varphi,\psi] + S^\text{G,V}_2[v^\text{T}_m] + S^\text{G,TT}_2[h^\text{TT}_{mn}] \eqend{,}
\end{split}
\end{equation}
where the scalar (S), vectorial (V) and tensorial (TT) parts have been separated, except for the nonlocal term $S_2^\text{K}$, which will be decomposed below. For the scalar part we have
\begin{equation}
\label{s0_scalar}
\begin{split}
S_0^\text{S}[\varphi, \psi] &= - \frac{3}{2} H^2 \int a^4 \varphi^2 \total^4 x + 2 \int H a^3 \varphi \laplace \psi \total^4 x \eqend{,} \\
S^\text{G,S}_2[\varphi, \psi] &= \frac{5}{2} \alpha H^4 \int a^4 \varphi^2 \total^4 x - 2 \left( 3 \alpha - \beta \right) H^2 \int H a^3 \varphi \laplace \psi \total^4 x \\
&\quad+ \alpha \int \left[ \frac{1}{2} \left( \laplace \varphi \right)^2 - 2 \left( \laplace \varphi \right) \laplace \psi' + 2 \left( \laplace \psi' \right)^2 \right] \ln a \total^4 x \\
&\quad+ \beta \int \bigg[ - \frac{7}{4} H^2 a^2 \varphi \laplace \varphi - \frac{3}{4} H^2 a^2 \left( \varphi' \right)^2 - \frac{1}{12} \left( \laplace \varphi \right)^2 - H a \varphi \laplace \psi'' + H a \left( \laplace \varphi \right) \laplace \psi \\
&\qquad\qquad+ \frac{1}{3} \left( \laplace \varphi \right) \laplace \psi' - 2 H^2 a^2 \left( \laplace \psi \right)^2 - \frac{1}{3} \left( \laplace \psi' \right)^2 \bigg] \total^4 x \eqend{.}
\end{split}
\end{equation}
where a prime is used to denote a derivative with respect to the conformal time $\eta$, $\laplace = \eta^{\mu\nu} \partial_\mu \partial_\nu$ and $\dalembert = \eta^{mn} \partial_m \partial_n = - \partial^2_\eta + \laplace$. On the other hand, for the vectorial part we have
\begin{equation}
\label{s0_vector}
\begin{split}
S_0^\text{V}[v^\text{T}_m] &= - \frac{1}{2} \int a^2 v_\text{T}^m \laplace v^\text{T}_m \total^4 x \eqend{,} \\
S_2^\text{G,V}[v^\text{T}_m] &= \frac{1}{2} \left( \alpha + 2 \beta \right) H^2 \int a^2 v_\text{T}^m \laplace v^\text{T}_m \total^4 x - \frac{3}{2} \alpha \int \left[ \left( \laplace v_\text{T}^m \right) \left( \laplace v^\text{T}_m \right) + v_\text{T}^{\prime m} \laplace v^{\text{T}\prime}_m \right] \ln a \total^4 x \eqend{,}
\end{split}
\end{equation}
and the tensorial part is given by
\begin{equation}
\label{tensor_S_G}
\begin{split}
S_0^\text{TT}[h^\text{TT}_{mn}] &= - \frac{1}{4} \int a^2 ( \partial_s h^\text{TT}_{mn} ) ( \partial^s h_\text{TT}^{mn} ) \total^4 x \eqend{,} \\
S_2^\text{G,TT}[h^\text{TT}_{mn}] &= - \frac{1}{4} (5 \alpha - 2 \beta) H^2 \int a^2 ( \partial_s h^\text{TT}_{mn} ) ( \partial^s h_\text{TT}^{mn} ) \total^4 x - \frac{3}{2} \alpha H^2 \int a^2 h^{\text{TT}\prime}_{mn} h_\text{TT}^{\prime mn} \total^4 x \\
&\quad+ \frac{3}{4} \alpha \int ( \dalembert h^\text{TT}_{mn} ) ( \dalembert h_\text{TT}^{mn} ) \ln a \total^4 x \eqend{.}
\end{split}
\end{equation}

The nonlocal part $S_2^\text{K}$ can also be decomposed into scalar, vectorial and tensorial parts when we take into account that the product of the two linearized Weyl tensors [integrated with a kernel as in equation~\eqref{effective_action_s2k}] can be decomposed in that way. In fact, for the scalar sector we have
\begin{equation}
\begin{split}
\left. C^{+abcd}(x) C^-_{abcd}(y) \right|_\text{S} &= \frac{\kappa^2}{3} \Big[ \left( \laplace \varphi^+(x) \right) \left( \laplace \varphi^-(y) \right) + 4 \left( \laplace \psi^{+\prime}(x) \right) \left( \laplace \psi^{-\prime}(y) \right) \\
&\qquad\qquad- 2 \left( \laplace \varphi^+(x) \right) \left( \laplace \psi^{-\prime}(y) \right) - 2 \left( \laplace \psi^{+\prime}(x) \right) \left( \laplace \varphi^-(y) \right) \Big] \eqend{,}
\end{split}
\end{equation}
for the vectorial sector we get
\begin{equation}
\left. C^{+abcd}(x) C^-_{abcd}(y) \right|_\text{V} = - \frac{\kappa^2}{2} \eta^{ab} \Big[ \left( \dalembert v^{+\text{T}}_a(x) \right) \left( \laplace v^{-\text{T}}_b(y) \right) + \left( \laplace v^{+\text{T}}_a(x) \right) \left( \dalembert v^{-\text{T}}_b(y) \right) \Big] \eqend{,}
\end{equation}
and the term for the tensorial sector is given by
\begin{equation}
\left. C^{+abcd}(x) C^-_{abcd}(y) \right|_\text{TT} = \frac{\kappa^2}{2} \left( \dalembert h^{+\text{TT}}_{ab}(x) \right) \eta^{ac} \eta^{bd} \left( \dalembert h^{-\text{TT}}_{cd}(y) \right) \eqend{.}
\end{equation}

Thus, up to the order at which we are working and which does not include nonlinear interactions for the metric perturbations, the effective action separates into three independent parts corresponding to the scalar, vectorial and tensorial perturbations of the metric. This in turn implies that the two-point function can also be written as the sum of scalar, vectorial and tensorial parts:
\begin{equation}
\label{2pt_meric}
G^{-+}_{abcd}(x,x') = G^{-+\text{S}}_{abcd}(x,x') + G^{-+\text{V}}_{abcd}(x,x') + G^{-+\text{TT}}_{abcd}(x,x') \eqend{,}
\end{equation}
where we have now restored the tensorial indices explicitly. In section \ref{2pf_vectorscalar} we calculate the one-loop contribution for the vectorial and scalar parts. The tensorial part, whose calculation is the most elaborate, was derived in \cite{frv2011a} and we merely restate the result in section \ref{2pf_tensor}.

The two-point function \eqref{2pt_meric} will be given in the exact gauge defined by equation~\eqref{metricpert_fixed_decomp}, and is not de Sitter-invariant. However, this should not be taken to conclude that the corresponding ``in'' vacuum state is not de Sitter-invariant, since it results from having chosen a gauge which is not de Sitter invariant. Indeed, the result for the gauge-invariant Riemann correlator obtained in section~\ref{curvature_riemann} will instead be manifestly de Sitter invariant.

\subsection{One-loop two-point function for the vectorial and scalar sector}
\label{2pf_vectorscalar}

The scalar and vectorial parts of the lowest-order CTP effective action \eqref{seff_powers_kappa} are given by equations~\eqref{s0_scalar} and \eqref{s0_vector}, respectively. Let us now Fourier transform the scalar and vectorial perturbations with respect to the spatial arguments to obtain
\begin{equation}
\label{lowest_order_effective_action_fourier}
S^\text{S}_0 + S^\text{V}_0 = \frac{1}{2} \int a^2 \vec{p}^2 \bar{v}_\text{T}^m \bar{v}^{\text{T}*}_m \frac{\total^3 p}{(2\mathpi)^3} \total \eta - \frac{3}{2} H^2 \int a^4 \bar{\varphi} \bar{\varphi}^* \frac{\total^3 p}{(2\mathpi)^3} \total \eta - 2 \int H a^3 \vec{p}^2 \bar{\varphi} \bar{\psi}^* \frac{\total^3 p}{(2\mathpi)^3} \total \eta \eqend{,}
\end{equation}
where all fields depend on $\eta$ and $\vec{p}$ and we have used bars to denote Fourier-transformed quantities with respect to the spatial coordinates. Introducing a right-handed orthonormal set of three vectors $\vec{e}_A(\vec{p})$ such that $\vec{p} = \abs{\vec{p}} \vec{e}_3$ and choosing that under reflection they satisfy
\begin{equation}
\vec{e}_3(-\vec{p}) = - \vec{e}_3(\vec{p}) \eqend{,} \qquad \vec{e}_1(-\vec{p}) = \vec{e}_2(\vec{p}) \eqend{,} \qquad \vec{e}_2(-\vec{p}) = \vec{e}_1(\vec{p}) \eqend{,}
\end{equation}
we can form the two (real) scalars
\begin{equation}
\bar{v}_+ = \frac{1}{2} ( \vec{e}_1^m + \vec{e}_2^m ) \bar{v}^\text{T}_m \eqend{,} \qquad \bar{v}_- = \frac{\mathi}{2} ( \vec{e}_1^m - \vec{e}_2^m ) \bar{v}^\text{T}_m \eqend{.}
\end{equation}
Taking into account the transversality condition $p^m \bar{v}^\text{T}_m = 0$, the vector $\bar{v}^\text{T}_m$ can be written as
\begin{equation}
\bar{v}_\text{T}^m = \bar{v}_+ \left( \vec{e}_1^m + \vec{e}_2^m \right) - \mathi \bar{v}_- \left( \vec{e}_1^m - \vec{e}_2^m \right) \eqend{.}
\end{equation}
The action \eqref{lowest_order_effective_action_fourier} then reads
\begin{equation}
\label{lowest_order_effective_action_fourier_scalar}
\begin{split}
S^\text{S}_0 + S^\text{V}_0 &= - \frac{3}{2} H^2 \int a^4 \bar{\varphi} \bar{\varphi}^* \frac{\total^3 p}{(2\mathpi)^3} \total \eta - 2 \int H a^3 \vec{p}^2 \bar{\varphi} \bar{\psi}^* \frac{\total^3 p}{(2\mathpi)^3} \total \eta \\
&\qquad+ \int a^2 \vec{p}^2 \left( \bar{v}_+ \bar{v}^*_+ + \bar{v}_- \bar{v}^*_- \right) \frac{\total^3 p}{(2\mathpi)^3} \total \eta \eqend{,}
\end{split}
\end{equation}
and the tree-level kinetic operator $A^0$ of equation~\eqref{a0_definition} has the following non-vanishing components:
\begin{equation}
\bar{A}^{0++}_{v_+v_+} = \bar{A}^{0++}_{v_-v_-} = \frac{2}{H^2 \eta^2} \vec{p}^2 \eqend{,} \quad \bar{A}^{0++}_{\varphi\varphi} = - \frac{3}{H^2 \eta^4} \eqend{,} \quad \bar{A}^{0++}_{\psi\psi} = 0 \eqend{,} \quad \bar{A}^{0++}_{\varphi\psi} = \bar{A}^0_{\psi\varphi} = \frac{2}{H^2 \eta^3} \vec{p}^2 \eqend{,}
\end{equation}
with $\bar{A}^{0--} = - \bar{A}^{0++}$ in every case.
We see that the scalar and vectorial perturbations satisfy elliptic equations (they become purely algebraic in spatial Fourier space) rather than hyperbolic ones, which reflects the fact they are not true dynamical degrees of freedom but constrained ones.
Analogously to the situation for the potential of the electromagnetic field in the Coulomb gauge, the CTP generating functional for the scalar and vectorial sector will still be given by equation~\eqref{ctp_generating_functional_free}, but with free ``propagators'' $G^0_{AB}$ proportional to $\delta (\eta-\eta')$ that result from solving the algebraic equation
\begin{equation}
\label{a0_op_components}
\begin{pmatrix} \bar{A}^{0++}(\eta, \vec{p}) & 0 \\ 0 & -\bar{A}^{0++}(\eta, \vec{p}) \end{pmatrix} \begin{pmatrix} \bar{G}^{0++}(\eta, \eta', \vec{p}) & \bar{G}^{0+-}(\eta, \eta', \vec{p}) \\ \bar{G}^{0-+}(\eta, \eta', \vec{p}) & \bar{G}^{0--}(\eta, \eta', \vec{p}) \end{pmatrix} = \unitmatrix \delta(\eta-\eta') \eqend{.}
\end{equation}
for their spatial Fourier transforms. By inverting the first matrix on the left-hand side of equation~\eqref{a0_op_components}, it follows that $\bar{G}^{0+-}(\eta, \eta', \vec{p}) = \bar{G}^{0-+}(\eta, \eta', \vec{p}) = 0$ as well as $\bar{G}^{0--}(\eta, \eta', \vec{p}) = - \bar{G}^{0++}(\eta, \eta', \vec{p})$, and for the latter ``propagator'' we have
\begin{equation}
\begin{split}
\bar{G}^{0++}_{v_+v_+} &= \bar{G}^{0++}_{v_-v_-} = \frac{H^2 \eta^2}{2 \vec{p}^2} \delta(\eta-\eta') \eqend{,} \\
\bar{G}^{0++}_{\varphi\psi} &= \bar{G}^{0++}_{\psi\varphi} = \frac{H^2 \eta^3}{2 \vec{p}^2} \delta(\eta-\eta') \eqend{,} \\
\bar{G}^{0++}_{\psi\psi} &= \frac{3 H^2 \eta^2}{4 (\vec{p}^2)^2} \delta(\eta-\eta') \eqend{,} \\
\bar{G}^{0++}_{\varphi\varphi} &= 0 \eqend{.}
\end{split}
\end{equation}
Using the decomposition \eqref{metricpert_fixed_decomp} for the metric perturbations, we readily obtain for the vector ``propagator''
\begin{equation}
\label{vector_propagator}
\bar{G}^{0++\text{V}}_{abcd}(\eta, \eta', \vec{p}) = 4 \delta^m_{(a} \delta^0_{b)} \delta^0_{(c} \delta^n_{d)} \left( \vec{e}^1_m \vec{e}^1_n + \vec{e}^2_m \vec{e}^2_n \right) \left( \bar{G}^{0++}_{v_+v_+} + \bar{G}^{0++}_{v_-v_-} \right) = 4 \delta^0_{(a} P_{b)(c} \delta^0_{d)} \frac{H^2 \eta^2}{\vec{p}^2} \delta(\eta-\eta') \eqend{,}
\end{equation}
where the projection tensor $P_{mn}$ is defined as
\begin{equation}
\label{projection_def}
P_{mn} = \eta_{mn} + \delta^0_m \delta^0_n - \frac{p_m p_n}{\vec{p}^2} \eqend{,}
\end{equation}
with $p^a = (0,\vec{p})$ in the spatially flat coordinate system that we have chosen. Similarly, for the scalar ``propagator'' we obtain
\begin{equation}
\label{scalar_propagator}
\begin{split}
\bar{G}^{0++\text{S}}_{abcd}(\eta, \eta', \vec{p}) &= 2 \mathi \left( \delta^0_{(a} p_{b)} \delta^0_c \delta^0_d - \delta^0_a \delta^0_b \delta^0_{(c} p_{d)} \right) \bar{G}^{0++}_{\phi\psi} + \delta^0_a \delta^0_b \delta^0_c \delta^0_d \bar{G}^{0++}_{\phi\phi} + 4 \delta^0_{(a} p_{b)} \delta^0_{(c} p_{d)} \bar{G}^{0++}_{\psi\psi} \\
&= \delta^0_{(a} \left[ \mathi \eta \left( p_{b)} \delta^0_{(c} - \delta^0_{b)} p_{(c} \right) + 3 \frac{p_{b)} p_{(c}}{\vec{p}^2} \right] \delta^0_{d)} \frac{H^2 \eta^2}{\vec{p}^2} \delta(\eta-\eta') \eqend{.}
\end{split}
\end{equation}

Because they are proportional to $\delta(\eta-\eta')$, when substituting the propagators \eqref{vector_propagator} and \eqref{scalar_propagator} into equation~\eqref{propagator_first_order_integral} to calculate the correlation function for the scalar and vectorial sector of the metric perturbations, the two time integrals can be trivially performed and the result is directly proportional to the corresponding component of the matrix $V_{AB}$ defined in equation~\eqref{perturbation_v_definition}.
(This is equivalent to solving the constraints corresponding to the Einstein equations for the scalar and vectorial perturbations, which become proportional, in Fourier space, to the stress tensor \cite{fprv2013}. Their correlator is then directly determined by the stress tensor correlator of the matter fields, which essentially coincides with $V_{AB}$.)
Furthermore, since the tree-level Wightman functions for the vectorial and scalar parts vanish, as seen above, the only parts of the matrix $V$ that contribute in this case to equation~\eqref{propagator_first_order_integral} are those with the index $M$ equal to $A$ and the index $N$ equal to $B$. Therefore, calculating the Wightman function at one-loop order, for which $AB = -+$, there will be no contribution from $V_{++}$ nor $V_{--}$ [which include all of $S^\text{G}_2$ in equation~\eqref{effective_action_decomp} as well as the terms with the two kernels $K^\pm$ in equation~\eqref{effective_action_s2k}]. Taking into account that $\bar{G}^{0--} = - \bar{G}^{0++}$, the vector contribution to the two-point function (the positive Wightman function) is then given by
\begin{equation}
\label{wightman_vector0}
\bar{G}^{-+\text{V}}_{abcd}(\eta, \eta', \vec{p}) = \kappa^2 \iint \bar{G}^{0++\text{V}}_{abmn}(\eta, \tau, \vec{p}) \bar{V}^{mnpq}_{-+}(\tau, \tau', \vec{p}) \bar{G}^{0++\text{V}}_{pqcd}(\tau', \eta', \vec{p}) \total \tau' \total \tau \eqend{,}
\end{equation}
where $V_{-+}$ is obtained by functionally differentiating the terms involving the kernel $K$ in equation~\eqref{effective_action_s2k}. After inserting the vector ``propagator'' \eqref{vector_propagator} and performing the time integrals, the result reads
\begin{equation}
\label{wightman_vector}
\bar{G}^{-+\text{V}}_{abcd}(\eta, \eta', \vec{p}) = 6 \mathi \alpha \kappa^2 \delta^0_{(a} P_{b)(c} \delta^0_{d)} H^4 \eta^2 (\eta')^2 \frac{1}{\vec{p}^2} \left( \partial^2_\eta + \vec{p}^2 \right) \bar{S}(\eta-\eta', \vec{p}) \eqend{,}
\end{equation}
where we defined the kernel
\begin{equation}
\label{def_kernel_s}
\bar{S}(\eta-\eta', \vec{p}) = - \mathi \mathe^{-\mathi \abs{\vec{p}} (\eta-\eta')} \mathcal{P} \frac{1}{\eta-\eta'} + \mathpi \delta(\eta-\eta') \eqend{.}
\end{equation}
This kernel is obtained (up to a proportionality factor) by the temporal Fourier transform of the kernel $K$ in equation~\eqref{kernels_def}\footnote{Details on such Fourier transforms can be found in appendix~D of \cite{frv2011a}.}. Here $\mathcal{P}$ denotes Hadamard's finite-part distribution \cite{frv2011a,schwartz}.

Proceeding in the same way, one finds that the scalar contribution is given by
\begin{equation}
\label{wightman_scalar}
\begin{split}
\bar{G}^{-+\text{S}}_{abcd}(\eta, \eta', \vec{p}) &= \frac{\mathi}{2} \alpha \kappa^2 H^4 \bigg[ 9 \delta^0_{(a} p_{b)} \delta^0_{(c} p_{d)} \frac{\eta^2 (\eta')^2}{(\vec{p}^2)^2} \partial^2_\eta + 3 \delta^0_{(a} p_{b)} \delta^0_{(c} p_{d)} \frac{\eta^2 (\eta')^2}{\vec{p}^2} (\eta-\eta') \partial_\eta \\
&\qquad- \delta^0_{(a} p_{b)} \delta^0_{(c} p_{d)} \eta^3 (\eta')^3 - 3 \mathi \delta^0_a \delta^0_b \delta^0_{(c} p_{d)} \frac{\eta^3 (\eta')^2}{\vec{p}^2} \partial^2_\eta + 3 \mathi \delta^0_{(a} p_{b)} \delta^0_c \delta^0_d \frac{\eta^2 (\eta')^3}{\vec{p}^2} \partial^2_\eta \\
&\qquad+ \mathi \delta^0_{(a} \left( \delta^0_{b)} p_{(c} + p_{b)} \delta^0_{(c} \right) \delta^0_{d)} \eta^3 (\eta')^3 \partial_\eta + \delta^0_a \delta^0_b \delta^0_c \delta^0_d \eta^3 (\eta')^3 \partial^2_\eta \bigg] \bar{S}(\eta-\eta', \vec{p}) \eqend{.}
\end{split}
\end{equation}

\subsection{One-loop two-point function for the tensorial sector}
\label{2pf_tensor}

The two-point function for the tensor perturbations is much more involved because in this case one needs to calculate the two time integrals in equation \eqref{propagator_first_order_integral}, but it was already obtained in \cite{frv2011a}. For the positive Wightman function the result can be written as
\begin{equation}
G^{-+\text{TT}}_{abcd}(x, x') = - \mathi \bra{\text{in}} h^\text{TT}_{ab}(x) h^\text{TT}_{cd}(x') \ket{\text{in}} = \int \bar{G}^{-+\text{TT}}_{abcd}(\eta, \eta', \vec{p}) \mathe^{\mathi \vec{p} (\vec{x}-\vec{x}')} \frac{\total^3 p}{(2\mathpi)^3} \eqend{,}
\end{equation}
with
\begin{equation}
\label{wightman_tensor}
\begin{split}
\bar{G}^{-+\text{TT}}_{abcd}(\eta, \eta', \vec{p}) &= - \mathi P_{abcd} \bigg[ f(\eta, \eta', \vec{p}) \left( 1 + 6 \alpha \kappa^2 H^2 \ln\left( \frac{\bar\mu}{H} \right) - \left( 5 \alpha - 2 \beta \right) \kappa^2 H^2 \right) \\
&\qquad\qquad+ \frac{3}{2} \alpha \kappa^2 H^4 g(\eta, \eta', \vec{p}) \bigg] + \bigo{\kappa^4} \eqend{,}
\end{split}
\end{equation}
where the polarization tensor $P_{abcd}$ is given by
\begin{equation}
\label{projector_def}
P_{abcd} = P_{ac} P_{bd} + P_{ad} P_{bc} - P_{ab} P_{cd} \eqend{,}
\end{equation}
in terms of the projection tensor $P_{ab}$ defined in equation~\eqref{projection_def}. The function $f(\eta, \eta', \vec{p})$ coincides with the tree-level result and reads
\begin{equation}
f(\eta, \eta', \abs{\vec{p}}) = \frac{H^2}{2 \abs{\vec{p}}^3} ( \abs{\vec{p}} \eta - \mathi ) ( \abs{\vec{p}} \eta' + \mathi ) \mathe^{- \mathi \abs{\vec{p}} (\eta-\eta')} \eqend{.}
\end{equation}
On the other hand, $g(\eta, \eta', \vec{p})$ can be written as
\begin{equation}
g(\eta, \eta', \vec{p}) = I_1(\eta, \eta', \vec{p}) - I_2(\eta, \eta', \vec{p}) - I^*_2(\eta', \eta, \vec{p}) + I_3(\eta, \eta', \vec{p}) - I_4(\eta, \eta', \vec{p}) + I_5(\eta, \eta', \vec{p}) \eqend{,}
\end{equation}
where the terms $I_k$ read
\begin{equation}
\begin{split}
I_1(\eta, \eta', \vec{p}) &= 2 \abs{\vec{p}}^{-1} \eta \eta' \mathe^{- \mathi \abs{\vec{p}} (\eta-\eta')} \\
I_2(\eta, \eta', \vec{p}) &= \abs{\vec{p}}^{-3} \mathe^{\mathi \abs{\vec{p}} (\eta+\eta')} \left( \abs{\vec{p}} \eta + \mathi \right) \left( \abs{\vec{p}} \eta' + \mathi \right) \left[ \Ein\left( - 2 \mathi \abs{\vec{p}} \eta \right) + \ln\left( 2 \mathi \abs{\vec{p}} \eta \right) + \gamma \right] \\
&\qquad+ \abs{\vec{p}}^{-3} \mathe^{- \mathi \abs{\vec{p}} (\eta-\eta')} \left( \abs{\vec{p}} \eta - \mathi \right) \left( \abs{\vec{p}} \eta' + \mathi \right) \ln\left( - 2 \abs{\vec{p}} \eta \right) \\
I_3(\eta, \eta', \vec{p}) &= \abs{\vec{p}}^{-3} \mathe^{- \mathi \abs{\vec{p}} (\eta-\eta')} \left( \abs{\vec{p}} \eta - \mathi \right) \left( \abs{\vec{p}} \eta' + \mathi \right) \left[ \ln\left( 2 \mathi \abs{\vec{p}} (\eta - \eta') \right) + \gamma \right] \\
I_4(\eta, \eta', \vec{p}) &= \abs{\vec{p}}^{-3} \mathe^{\mathi \abs{\vec{p}} (\eta-\eta')} \left( \abs{\vec{p}} \eta + \mathi \right) \left( \abs{\vec{p}} \eta' - \mathi \right) \left[ \Ein\left( - 2 \mathi \abs{\vec{p}} (\eta-\eta') \right) + \ln\left( 2 \mathi \abs{\vec{p}} (\eta - \eta') \right) + \gamma \right] \\
I_5(\eta, \eta', \vec{p}) &= \eta^2 (\eta')^2 \bar{S}(\eta-\eta', \vec{p}) \eqend{.}
\end{split}
\end{equation}
The kernel $\bar{S}$ was defined in equation~\eqref{def_kernel_s}, $\gamma$ is the Euler-Mascheroni constant and the special function $\Ein$ is defined in appendix \ref{appendix_special}. Note that this expression is invariant under the simultaneous rescaling $\{ \vec{p} \to \lambda^{-1} \vec{p}; \eta, \eta' \to \lambda \eta, \lambda \eta' \}$ except for an inverse cubic prefactor, which is a necessary (but not sufficient) condition for de Sitter invariance. It should also be noted that this result is valid for the choice of renormalization scale $\bar\mu$ such that $a_1^\text{ren}(\bar\mu) = 0$. As explained in section~\ref{2pf_action}, for arbitrary values of the renormalization scale $\mu$ one needs to make the replacement $\alpha \ln\bar\mu \to N a_1^\text{ren}(\mu) + \alpha \ln \mu$.

\section{Maximally symmetric bitensors}
\label{bitensors}

Any de Sitter-invariant two-point correlation function can be expressed in terms of maximally symmetric bitensors \cite{allenjacobson86}. Here we will review the suitable formalism adapted to our problem. Four-dimensional de Sitter spacetime may be defined as a submanifold (an hyperboloid) embedded in five-dimensional Minkowski spacetime whose points satisfy the equation $\eta_{AB} X^A(x) X^B(x) = H^{-2}$ ($A, B = 0, \ldots, 4$) where $H$ is a parameter with dimensions of mass, $\eta_{AB}$ is the Minkowski metric and $X^A(x)$ is the position vector in Minkowski spacetime corresponding to the point $x$ of de Sitter spacetime. De Sitter spacetime then inherits all the symmetries of the embedding Minkowski spacetime which leave the de Sitter submanifold unchanged. The basic de Sitter-invariant biscalar is defined as $Z(x,x') = H^2 \eta_{AB} X^A(x) X^B(x')$. It is related to the Minkowski interval between two points $X^A(x)$ and $X^B(x')$ of the submanifold by
\begin{equation}
\label{minkowski_z}
\eta_{AB} (X^A(x) - X^A(x')) (X^B(x) - X^B(x')) = 2 H^{-2} ( 1 - Z(x,x') ) \eqend{.}
\end{equation}
Thus, since the interval between pairs of points on the submanifold can take any real value, we also have $Z(x,x') \in (-\infty,\infty)$. Additionally, we see that $Z(x,x') = 1$ when $x'$ is on the light cone of $x$, $Z(x,x') > 1$ when the two points are timelike separated and $Z(x,x') < 1$ when they are spacelike separated. The geodesic distance $\mu(x,x')$ along the shortest geodesic connecting the points $x$ and $x'$ is related to the biscalar $Z(x,x')$ by \cite{allen85}
\begin{equation}
\label{relation_mu_z}
Z(x,x') = \cos \left( H \mu(x, x') \right) \eqend{.}
\end{equation}
For points which are not connected by a geodesic [those for which $Z(x,x') < -1$], $\mu$ can still be defined by a suitable analytic continuation from \eqref{relation_mu_z}.
This can also be done for the bitensors defined below and the result has a simple and appealing geometric interpretation as shown in \cite{perez-nadal10}.

There exists a complete set of de Sitter-invariant bitensors from which all maximally symmetric bitensors can be constructed \cite{allenjacobson86}. This set is given by the metric $\hat{g}_{ab}(x)$, the geodesic distance $\mu(x,x')$, the unit tangent vectors to the geodesic at the points $x$ and $x'$ defined by
\begin{equation}
n_a(x,x') = \partial_a \mu(x,x') \eqend{,} \qquad n_{a'}(x,x') = \partial_{a'} \mu(x,x') \eqend{,}
\end{equation}
and the propagator $\hat{g}_{ab'}(x,x')$ which parallely transports a vector $w_a$ from $x'$ to $x$ along the geodesic connecting them: $w_a(x) = \hat{g}_a^{b'}(x,x') w_{b'}(x')$. (Remember that we use a hat over a symbol to refer to quantities which are defined in the background de Sitter space.) A theorem by Allen and Jacobson~\cite{allenjacobson86} establishes that any maximally symmetric bitensor is a linear combination of products of those basic bitensors with coefficients that only depend on the geodesic distance $\mu(x,x')$. By inverting equation~\eqref{relation_mu_z}, we can write the complete set of bitensors described above in terms of covariant derivatives of $Z(x,x')$ as follows:
\begin{equation}
\label{relation_z_n}
\begin{gathered}
n_a = - \frac{1}{H \sqrt{1-Z^2}} \hat{\nabla}_a Z \eqend{,} \qquad n_{a'} = - \frac{1}{H \sqrt{1-Z^2}} \hat{\nabla}_{a'} Z \eqend{,} \\
H^2 \hat{g}_{ab'} = \hat{\nabla}_a \hat{\nabla}_{b'} Z - \frac{1}{1 + Z} ( \hat{\nabla}_a Z ) ( \hat{\nabla}_{b'} Z ) \eqend{.}
\end{gathered}
\end{equation}
In addition, as an illustration of the above theorem, we have \cite{allenjacobson86} that
\begin{equation}
\label{normal_derivative}
\hat{\nabla}_a n_b = \frac{Z}{\sqrt{1-Z^2}} H \left( \hat{g}_{ab} - n_a n_b \right) \eqend{,}
\end{equation}
from which we can calculate the second derivative of $Z$ at one point as
\begin{equation}
\label{z_second_derivative}
\hat{\nabla}_a \hat{\nabla}_b Z = - \sqrt{1-Z^2} \, H \hat{\nabla}_a n_b - Z H^2 n_a n_b = - Z H^2 \hat{g}_{ab} \eqend{.}
\end{equation}
In many calculations it is more convenient to use the set provided by covariant derivatives of $Z$, because the application of a covariant derivative directly gives another member of this set [with the only exception shown in~\eqref{z_second_derivative}, which has a very simple structure]. In contrast, if one uses the set introduced by Allen and Jacobson, one encounters more or less complicated prefactors as in equation~\eqref{normal_derivative}.

In the conformally flat coordinate system of the Poincaré patch, we can use equation~\eqref{minkowski_z} together with the embedding map $X^A(x)$ for these coordinates,
\begin{equation}
\label{flat_embedding}
\begin{split}
X^0 &= - \frac{1+\vec{x}^2}{2\,\eta} + \frac{\eta}{2} \eqend{,} \nonumber \\
X^i &= - \frac{x^i}{\eta} \eqend{,}  \quad i=1,2,3 \nonumber \\
X^4 &= \frac{1-\vec{x}^2}{2\,\eta} + \frac{\eta}{2} \eqend{,}
\end{split}
\end{equation}
to obtain
\begin{equation}
\label{z_flat_coordinates}
Z(x, x') = 1 - \frac{(x-x')^2}{2 \eta \eta'} = 1 - \frac{(\vec{x}-\vec{x}')^2 - (\eta-\eta')^2}{2 \eta \eta'} \eqend{,}
\end{equation}
from which we can derive
\begin{equation}
\label{z_flat_coordinates_deriv}
\begin{aligned}
\partial_\eta Z &= \frac{1}{\eta'} - \frac{Z}{\eta} \eqend{,} &\qquad \partial_{\eta'} Z &= \frac{1}{\eta} - \frac{Z}{\eta'} \eqend{,} \\
\partial_\alpha Z &= - \frac{\eta_{\alpha\beta} (\vec{x} - \vec{x}')^\beta}{\eta \eta'} \eqend{,} &\qquad \partial_{\alpha'} Z &= \frac{\eta_{\alpha\beta} (\vec{x} - \vec{x}')^\beta}{\eta \eta'} \eqend{.}
\end{aligned}
\end{equation}
In the next section we will also need contractions of the derivatives of $Z$. To shorten the notation, from now on we will use a semicolon to denote the covariant derivative $\hat{\nabla}_a$ with respect to the background de Sitter metric, i.e. $Z_{;a} = \hat{\nabla}_a Z$. Using equations~\eqref{relation_z_n} as well as the normalization $n_a n^a = 1$ and $n_{a'} = - \hat{g}_{a'b} n^b$, which implies $Z_{;a'} = - \hat{g}_{a'b} Z^{;b}$, we get
\begin{equation}
\label{derivatives_z_contracted}
\begin{aligned}
Z_{;a} Z^{;a} &= H^2 (1-Z^2) \eqend{,} &\qquad Z_{;a'} Z^{;a'} &= H^2 (1-Z^2) \eqend{,} \\
Z_{;ab'} Z^{;a} &= - H^2 Z Z_{;b'} \eqend{,} &\qquad Z_{;ab'} Z^{;b'} &= - H^2 Z Z_{;a} \eqend{,} \\
Z_{;ab'} Z^{;ac'} &= H^4 \delta_{b'}^{c'} - H^2 Z_{;b'} Z^{;c'} \eqend{,} &\qquad Z_{;ab'} Z^{;cb'} &= H^4 \delta_a^c - H^2 Z_{;a} Z^{;c} \eqend{.}
\end{aligned}
\end{equation}
Of course, this can also be directly calculated from expression~\eqref{z_flat_coordinates} for $Z$ in the spatially flat coordinates.

\section{Curvature tensor correlators}
\label{curvature}

In this section we compute the two-point correlator for the Riemann tensor and express it in terms of de Sitter-invariant bitensors. The linearized Riemann tensor $\tilde{R}^{ab}{}_{cd}$ of the perturbed de Sitter spacetime can be written in terms of the metric perturbations and its derivatives as
\begin{equation}
\label{riemann_perturbed}
\begin{split}
{\tilde{R}^{ab}}{}_{cd} &= 2 H^2 \delta^a_{[c} \delta_{d]}^b \left( 1 + \kappa h_{00} \right) + 2 \kappa H^2 \eta^2 \eta^{m[a} \eta^{b]n} \partial_n \partial_{[c} h_{d]m} \\
&\quad+ 2 \kappa H^2 \eta \delta^m_{[c} \delta_{d]}^{[b} \eta^{a]n} \left( 2 \partial_{(m} h_{n)0} - \partial_0 h_{mn} \right) + \bigo{\kappa^2} \\
&= 2 H^2 \delta^a_{[c} \delta_{d]}^b + \mathcal{T}^{ab}{}_{cd}{}^{mn} h_{mn} + \bigo{\kappa^2} \eqend{,}
\end{split}
\end{equation}
where we have used appendices \ref{appendix_metric} and \ref{appendix_conformal} for the perturbative expansion in $\kappa$ in the first equality, and we define the differential operator $\mathcal{T}$ in the second equality. One can explicitly check the gauge invariance of this expression using the gauge transformation \eqref{gauge_trafo}. Note that it is only with the given index position that gauge invariance holds, because then the Lie derivative of the unperturbed tensor (evaluated for the background metric $\hat{g}_{ab}$) with respect to an arbitrary vector field $w^a$ vanishes:
\begin{equation}
\mathcal{L}_w {\hat{R}^{ab}}{}_{cd} = \mathcal{L}_w \left( 2 H^2 \delta^a_{[c} \delta_{d]}^b \right) = 0 \eqend{,}
\end{equation}
and this corresponds to the diffeomorphisms generated by $w^a$ at linear order.
To calculate the Riemann tensor correlator, we apply the differential operator $\mathcal{T}$ defined in equation~\eqref{riemann_perturbed} to each argument of the two-point function of the metric perturbations given in section~\ref{2pf}. If we consider only the connected correlator
\begin{equation}
\expect{A(x) B(x')} \equiv \bra{\text{in}} A(x) B(x') \ket{\text{in}} - \bra{\text{in}} A(x) \ket{\text{in}} \bra{\text{in}} B(x') \ket{\text{in}} \eqend{,}
\end{equation}
the terms of order $\kappa^2$ in the Riemann tensor, which would give a contribution when combined with the background value of the Riemann tensor, cancel out. Consequently, to lowest order we have
\begin{equation}
\label{riemann_correlator}
\expect{{\tilde{R}^{ab}}{}_{cd}(x) {\tilde{R}^{m'n'}}{}_{p'q'}(x')} = 4 \kappa^2 H^4 \hat{g}^{ak} \hat{g}^{bl} \hat{g}^{m's'} \hat{g}^{n't'} U_{[kl][cd][s't'][p'q']}(x, x') \eqend{,}
\end{equation}
where the bitensor $U$ is defined by its Fourier transform
\begin{equation}
\label{riemann_correlator_in_t}
\bar{U}_{abcdm'n'p'q'}(\eta, \eta', \vec{p}) = \frac{1}{H^8 \eta^4 (\eta')^4} \sum_{k,l=0}^2 T^{(k)}{}_{abcd}{}^{kl}(\eta) T^{(l)*}{}_{m'n'p'q'}{}^{t'u'}(\eta') \Big( \mathi \bar{G}^{-+}_{klt'u'}(\eta, \eta', \vec{p}) \Big)
\end{equation}
with the differential operators
\begin{equation}
\begin{split}
T^{(0)}{}_{abcd}{}^{kl}(\eta) &= \left( \eta_{ac} \eta_{bd} \delta_0^k \delta_0^l + 2 \mathi \eta \eta_{ac} p_{(b} \delta_{d)}^k \delta_0^l + \eta^2 p_a p_c \delta_b^k \delta_d^l \right) \\
T^{(1)}{}_{abcd}{}^{kl}(\eta) &= \left( - \eta \eta_{ac} \delta_b^k \delta_d^l + 2 \eta \eta_{ac} \delta^0_{(b} \delta_{d)}^k \delta_0^l - 2 \mathi \eta^2 p_{(a} \delta^0_{c)} \delta_b^k \delta_d^l \right) \partial_\eta \\
T^{(2)}{}_{abcd}{}^{kl}(\eta) &= - \eta^2 \delta^0_a \delta^0_c \delta_b^k \delta_d^l \, \partial^2_\eta \eqend{,}
\end{split}
\end{equation}
where the superscript in parentheses indicates the number of time derivatives involved.

\subsection{The Riemann correlator}
\label{curvature_riemann}

After calculating the bitensor $\bar{U}$ defined in equation~\eqref{riemann_correlator_in_t}, we have to invert the Fourier transform and write in terms of maximally symmetric bitensors the resulting tensor structure, which will consist of linear combinations of $\eta_{ab}$, $\delta^0_a$ and $r_a = \eta_{ab} r^b$, where $r^a = (0,\vec{r})$ with $\vec{r} = \vec{x} - \vec{x}'$. The coefficient functions, which will be functions of $\eta$, $\eta'$ and $\vec{r}$, also have to be written in terms of the de Sitter-invariant biscalar $Z(x,x')$. All the necessary substitutions can be deduced from expression~\eqref{z_flat_coordinates} for $Z(x,x')$ and its derivatives~\eqref{z_flat_coordinates_deriv} in our spatially flat coordinate system. Explicitly, we have
\begin{equation}
\label{bitensors_replacement}
\begin{aligned}
\eta_{ab} &\to H^2 \eta^2 \hat{g}_{ab} \eqend{,} & r_a &\to - \eta \eta' Z_{;a} + \delta^0_a \left( \eta - \eta' Z \right) \eqend{,} \\
\eta_{a'b'} &\to H^2 (\eta')^2 \hat{g}_{a'b'} \eqend{,} & r_{a'} &\to \eta \eta' Z_{;a'} - \delta^0_{a'} \left( \eta' - \eta Z \right) \eqend{,} \\
\eta_{ab'} &\to \eta \eta' Z_{;ab'} + \eta \delta^0_{b'} Z_{;a} + \eta' \delta^0_a Z_{;b'} + \delta^0_a \delta^0_{b'} (Z-1) \eqend{.} \hspace{-15em} &&
\end{aligned}
\end{equation}
Technical details for the calculation of the Fourier transforms are provided in appendix~\ref{appendix_fts}.

It turns out that after a long calculation, computing the Fourier transforms and implementing the substitutions described above, all the terms which are not de Sitter-invariant cancel out. We are finally left with a de Sitter-invariant result which can be expressed as
\begin{equation}
\label{riemann_correlator_pos}
\begin{split}
&\expect{{\tilde{R}^{ab}}{}_{cd}(x) {\tilde{R}^{m'n'}}{}_{p'q'}(x')} = \\
&\quad\frac{4 \kappa^2 H^6}{\mathpi^2} \sum_{k=1}^9 {}^{(k)}\mathcal{R}^{[ab]}{}_{[cd]}{}^{[m'n']}{}_{[p'q']}(Z(x,x')) \distlim_{\epsilon \to 0^+} \bigg[ \frac{3}{2} \alpha \kappa^2 H^2 \mathcal{R}^{(1,k)}(Z(x,x') - \mathi\sgneps) \\
&\qquad\quad+ \left( 1 + 6 \alpha \kappa^2 H^2 \left( \ln\left( \frac{\bar{\mu}}{H} \right) + \gamma \right) - \left( 5 \alpha - 2 \beta \right) \kappa^2 H^2 \right) \mathcal{R}^{(0,k)}(Z(x,x') - \mathi\sgneps) \bigg] \eqend{.}
\end{split}
\end{equation}
Here $\sgneps = +\epsilon$ whenever $x$ lies within the future light cone of $x'$, while $\sgneps = -\epsilon$ when $x$ lies within the past light cone of $x'$. If both points are spacelike separated, $Z(x,x') < 1$ and the sign of $\epsilon$ is irrelevant. The limit $\epsilon \to 0^+$ is to be taken in the sense of distributions, i.e., after smearing with a test function, which we indicate with the notation ``$\distlim$''. Furthermore, ${}^{(k)}\mathcal{R}$ are a set of nine 8-index bitensors constructed from the metric and covariant derivatives of $Z$, which have the appropriate symmetries [when antisymmetrized in each pair of indices as indicated in equation \eqref{riemann_correlator_pos}]:
\begin{equation}
\label{basisbitensors_r}
\begin{split}
{}^{(1)}\mathcal{R}_{abcdm'n'p'q'}(Z) &= \hat{g}_{ac} \, \hat{g}_{bd} \, \hat{g}_{m'p'} \, \hat{g}_{n'q'} \\
{}^{(2)}\mathcal{R}_{abcdm'n'p'q'}(Z) &= H^{-2} \hat{g}_{ac} \, \hat{g}_{m'p'} \left( \hat{g}_{bd} \, Z_{;n'} Z_{;q'} + Z_{;b} Z_{;d} \, \hat{g}_{n'q'} \right) \\
{}^{(3)}\mathcal{R}_{abcdm'n'p'q'}(Z) &= H^{-4} \hat{g}_{ac} \, \hat{g}_{m'p'} Z_{;b} Z_{;d} Z_{;n'} Z_{;q'} \\
{}^{(4)}\mathcal{R}_{abcdm'n'p'q'}(Z) &= 4 H^{-4} \hat{g}_{ac} \, \hat{g}_{m'p'} Z_{;(b} Z_{;d)(n'} Z_{;q')} \\
{}^{(5)}\mathcal{R}_{abcdm'n'p'q'}(Z) &= 2 H^{-4} \hat{g}_{ac} \, \hat{g}_{m'p'} Z_{;b(n'} Z_{;q')d} \\
{}^{(6)}\mathcal{R}_{abcdm'n'p'q'}(Z) &= 2 H^{-6} \left( \hat{g}_{ac} \, Z_{;m'} Z_{;p'} + Z_{;a} Z_{;c} \, \hat{g}_{m'p'} \right) Z_{;b(n'} Z_{;q')d} \\
{}^{(7)}\mathcal{R}_{abcdm'n'p'q'}(Z) &= 2 H^{-8} Z_{;a} Z_{;c} Z_{;m'} Z_{;p'} \, Z_{;b(n'} Z_{;q')d} \\
{}^{(8)}\mathcal{R}_{abcdm'n'p'q'}(Z) &= 8 H^{-8} Z_{;(a} Z_{;c)(m'} Z_{;p')} \, Z_{;b(n'} Z_{;q')d} \\
{}^{(9)}\mathcal{R}_{abcdm'n'p'q'}(Z) &= 4 H^{-8} Z_{;a(m'} Z_{;p')c} \, Z_{;b(n'} Z_{;q')d} \eqend{.}
\end{split}
\end{equation}
The coefficients $\mathcal{R}^{(0,k)}(Z)$ and $\mathcal{R}^{(1,k)}(Z)$ are functions of $Z$ and are, therefore, manifestly de Sitter-invariant biscalars. Since they are rather long, we give their explicit expressions in appendix~\ref{appendix_riemann2pf}. An important feature of those coefficients is that they are well-defined distributions. Regarded as functions, they are only singular as $Z \to 1$, i.e., when $x$ and $x'$ are null separated, which is also the case for the two-point function of the matter field $\phi$.

We should emphasize that this is a rather remarkable and nontrivial result. De Sitter invariance has not been assumed but derived from the the two-point metric correlations computed in the Poincaré patch, which are not only gauge-dependent but also not manifestly de Sitter invariant. The only de Sitter-invariant input is the selection of a de Sitter-invariant initial vacuum. This result should also be seen as a nontrivial check of the contributions to the two-point correlation functions of the metric perturbations, given by~\eqref{wightman_vector}, \eqref{wightman_scalar} and, more importantly, \eqref{wightman_tensor}.

As explained in section~\ref{bitensors}, we may as well use the complete set of basis bitensors which involve $n_a$, $n_{a'}$ and $\hat{g}_{ab'}$ (we do not write explicitly the dependence of those de Sitter-invariant bitensors on $x$ and $x'$). In this basis the Riemann correlator reads
\begin{equation}
\label{riemann_correlator_nn}
\begin{split}
&\expect{{\tilde{R}^{ab}}{}_{cd}(x) {\tilde{R}^{m'n'}}{}_{p'q'}(x')} = \\
&\quad\frac{4 \kappa^2 H^6}{\mathpi^2} \sum_{k=1}^9 {}^{(k)}\mathcal{N}^{[ab]}{}_{[cd]}{}^{[m'n']}{}_{[p'q']} \distlim_{\epsilon \to 0^+} \bigg[ \frac{3}{2} \alpha \kappa^2 H^2 \mathcal{S}^{(1,k)}(Z(x,x') - \mathi\sgneps) \\
&\qquad\quad+ \left( 1 + 6 \alpha \kappa^2 H^2 \left( \ln\left( \frac{\bar{\mu}}{H} \right) + \gamma \right) - \left( 5 \alpha - 2 \beta \right) \kappa^2 H^2 \right) \mathcal{S}^{(0,k)}(Z(x,x') - \mathi\sgneps) \bigg] \eqend{,}
\end{split}
\end{equation}
where ${}^{(k)}\mathcal{N}$ are a set of nine 8-index de Sitter-invariant bitensors constructed with the basis of bitensors introduced by Allen and Jacobson, and which have also the appropriate symmetries (when antisymmetrized in each pair of indices):
\begin{equation}
\label{basisbitensors_n}
\begin{split}
{}^{(1)}\mathcal{N}_{abcdm'n'p'q'} &= \hat{g}_{ac} \, \hat{g}_{bd} \, \hat{g}_{m'p'} \, \hat{g}_{n'q'} \\
{}^{(2)}\mathcal{N}_{abcdm'n'p'q'} &= \hat{g}_{ac} \, \hat{g}_{m'p'} \left( \hat{g}_{bd} \, n_{n'} n_{q'} + n_b n_d \, \hat{g}_{n'q'} \right) \\
{}^{(3)}\mathcal{N}_{abcdm'n'p'q'} &= \hat{g}_{ac} \, \hat{g}_{m'p'} \, n_b n_d n_{n'} n_{q'} \\
{}^{(4)}\mathcal{N}_{abcdm'n'p'q'} &= 4 \hat{g}_{ac} \, \hat{g}_{m'p'} \, n_{(b} \hat{g}_{d)(n'} n_{q')} \\
{}^{(5)}\mathcal{N}_{abcdm'n'p'q'} &= 2 \hat{g}_{ac} \, \hat{g}_{m'p'} \, \hat{g}_{b(n'} \hat{g}_{q')d} \\
{}^{(6)}\mathcal{N}_{abcdm'n'p'q'} &= 2 \left( \hat{g}_{ac} \, n_{m'} n_{p'} + n_a n_c \, \hat{g}_{m'p'} \right) \hat{g}_{b(n'} \hat{g}_{q')d} \\
{}^{(7)}\mathcal{N}_{abcdm'n'p'q'} &= 2 n_a n_c n_{m'} n_{p'} \, \hat{g}_{b(n'} \hat{g}_{q')d} \\
{}^{(8)}\mathcal{N}_{abcdm'n'p'q'} &= 8 n_{(a} \hat{g}_{c)(m'} n_{p')} \, \hat{g}_{b(n'} \hat{g}_{q')d} \\
{}^{(9)}\mathcal{N}_{abcdm'n'p'q'} &= 4 \hat{g}_{a(m'} \hat{g}_{p')c} \, \hat{g}_{b(n'} \hat{g}_{q')d} \eqend{.}
\end{split}
\end{equation}
The explicit expressions for the coefficients $\mathcal{S}^{(0,k)}(Z)$ and $\mathcal{S}^{(1,k)}(Z)$ are given in appendix~\ref{appendix_riemann2pf}.

Our result for the two-point function of the Riemann tensor~\eqref{riemann_correlator_pos} or~\eqref{riemann_correlator_nn} is manifestly de Sitter invariant and, in addition to the index symmetries mentioned above, they obey both the first and second Bianchi identities as expected\footnote{Note that for the Bianchi identities to be satisfied, 
including the (distributional) kernel $\bar{S}$ in the two-point function of the tensorial metric perturbations~\eqref{wightman_tensor} is essential.}. In order to analyze its structure, it is better to decompose the Riemann correlator in terms of its Weyl and Ricci parts:
\begin{equation}
\label{riemann_correlator_decomp}
\begin{split}
\expect{{\tilde{R}^{ab}}{}_{cd}(x) {\tilde{R}^{m'n'}}{}_{p'q'}(x')} &= \expect{{\tilde{C}^{ab}}{}_{cd}(x) {\tilde{C}^{m'n'}}{}_{p'q'}(x')} + 4 \delta^{[a}_{[c} \expect{{\tilde{R}^{b]}}{}_{d]}(x) {\tilde{R}^{[m'}}{}_{[p'}(x')} \delta^{n']}_{q']} \\
&\quad+ 2 \delta^{[a}_{[c} \expect{{\tilde{R}^{b]}}{}_{d]}(x) {\tilde{C}^{m'n'}}{}_{p'q'}(x')} + 2 \expect{{\tilde{C}^{ab}}{}_{cd}(x) {\tilde{R}^{[m'}}{}_{[p'}(x')} \delta^{n']}_{q']} \eqend{,}
\end{split}
\end{equation}
where the correlation functions on the right-hand side are obtained from the Riemann correlator by contracting indices (for the Ricci tensor) and by subtracting traces from the Riemann correlator (for the Weyl tensor). In the next two subsections, we discuss in more detail the main properties of these correlators. From the previous decomposition it also follows that the scalar correlators defined as
\begin{equation}
\label{quadrat_correlator_curvature}
\begin{aligned}
\expect{\mathrm{\tilde{R}ie}{}^2(x,x')} &\equiv \hat{g}_{am'} \hat{g}_{bn'} \hat{g}^{cp'} \hat{g}^{dq'} \expect{{\tilde{R}^{ab}}{}_{cd}(x) {\tilde{R}^{m'n'}}{}_{p'q'}(x')} \eqend{,} \\
\expect{\mathrm{\tilde{W}eyl}{}^2(x,x')} &\equiv \hat{g}_{am'} \hat{g}_{bn'} \hat{g}^{cp'} \hat{g}^{dq'} \expect{{\tilde{C}^{ab}}{}_{cd}(x) {\tilde{C}^{m'n'}}{}_{p'q'}(x')} \eqend{,} \\
\expect{\mathrm{\tilde{R}ic}{}^2(x,x')} &\equiv \hat{g}_{bn'} \hat{g}^{dq'} \expect{{\tilde{R}^{b}}{}_{d}(x) {\tilde{R}^{n'}}{}_{q'}(x')} \eqend{,}
\end{aligned}
\end{equation}
are related by the following identity:
\begin{equation}
\label{quadrat_correlator_riemann}
\expect{\mathrm{\tilde{R}ie}{}^2(x,x')} = \expect{\mathrm{\tilde{W}eyl}{}^2(x,x')} + 2 \expect{\mathrm{\tilde{R}ic}{}^2(x,x')} \eqend{.}
\end{equation}
Hereafter the dependence of $Z$ on $x$ and $x'$ should always be implicitly understood.

Note that even though the correlation functions of the metric perturbations were calculated in the Poincaré patch (this was essential in order to select a de Sitter-invariant vacuum for the interacting theory), when written in one of the de Sitter-invariant bases, the results for the Riemann correlator can be extended to the full de Sitter spacetime.

\subsection{The Weyl correlator}
\label{curvature_weyl}

The Weyl tensor in four dimensions can be calculated by subtracting traces from the Riemann tensor as follows:
\begin{equation}
\label{weyl_projection_riemann}
C^{ab}{}_{cd} = \left( \delta^a_k \delta^b_l \delta^s_c \delta^t_d + 2 \delta_k^s \delta^{[a}_l \delta^{b]}_{[c} \delta^t_{d]} + \frac{1}{3} \delta^{[a}_{[c} \delta^{b]}_{d]} \delta_k^s \delta_l^t \right) R^{kl}{}_{st} \eqend{.}
\end{equation}
We can then calculate its correlator very easily from the Riemann correlator~\eqref{riemann_correlator_pos}. However, because of the tracelessness of the Weyl tensor, the corresponding coefficients for the bitensor set \eqref{basisbitensors_r} are not independent since those basic bitensors are not traceless. In fact, there are only three combinations of invariant bitensors which have vanishing trace on any contraction. These are
\begin{equation}
\label{basisbitensors_weyl}
\begin{split}
{}^{(1)}\mathcal{C}_{[ab][cd][m'n'][p'q']}(Z) &= \left[ - 2 Z {}^{(1)}\mathcal{R} - 3 {}^{(4)}\mathcal{R} + 6 Z {}^{(5)}\mathcal{R} + 2 {}^{(8)}\mathcal{R} - 2 Z {}^{(9)}\mathcal{R} \right]_{[ab][cd][m'n'][p'q']} (Z) \eqend{,}\\
{}^{(2)}\mathcal{C}_{[ab][cd][m'n'][p'q']}(Z) &= \Big[ - (5-Z^2) {}^{(1)}\mathcal{R} + 6 {}^{(2)}\mathcal{R} + 12 {}^{(5)}\mathcal{R} - 6 {}^{(6)}\mathcal{R} - 6 {}^{(7)}\mathcal{R} \\
&\qquad\qquad\qquad\qquad\qquad\quad\,+ 2 Z {}^{(8)}\mathcal{R} - (3+Z^2) {}^{(9)}\mathcal{R} \Big]_{[ab][cd][m'n'][p'q']}(Z) \eqend{,}\\
{}^{(3)}\mathcal{C}_{[ab][cd][m'n'][p'q']}(Z) &= \left[ (1-Z^2) \left[ {}^{(1)}\mathcal{R} - 3 {}^{(5)}\mathcal{R} + {}^{(9)}\mathcal{R} \right] - 6 {}^{(3)}\mathcal{R} + 6 {}^{(7)}\mathcal{R} \right]_{[ab][cd][m'n'][p'q']}(Z) \eqend{.}
\end{split}
\end{equation}
We, therefore, obtain the following result for the correlator of the Weyl tensor for perturbations around de Sitter spacetime including radiative corrections from conformal matter fields up to order $\kappa^4$ (one-loop order for free matter fields):
\begin{equation}
\label{weyl_correlator}
\begin{split}
\expect{{\tilde{C}^{ab}}{}_{cd}(x) {\tilde{C}^{m'n'}}{}_{p'q'}(x')} &= \frac{4 \kappa^2 H^6}{\mathpi^2} \sum_{k=1}^3 {}^{(k)}\mathcal{C}^{[ab]}{}_{[cd]}{}^{[m'n']}{}_{[p'q']}(Z) \distlim_{\epsilon \to 0^+} \bigg[ \frac{\alpha}{4} \kappa^2 H^2 \mathcal{C}^{(1,k)}(Z-\mathi\sgneps) \\
&+ \left( 1 + 6 \alpha \kappa^2 H^2 \left( \ln\left( \frac{\bar{\mu}}{H} \right) + \gamma \right) - \left( 5 \alpha - 2 \beta \right) \kappa^2 H^2 \right) \mathcal{C}^{(0,k)}(Z-\mathi\sgneps) \bigg] \eqend{,}
\end{split}
\end{equation}
where the coefficients $\mathcal{C}^{(0,k)}(Z)$ and $\mathcal{C}^{(1,k)}(Z)$ are given explicitly by
\begin{equation}
\begin{split}
\mathcal{C}^{(0,1)}(Z) &= \frac{1}{4} (2-Z) (1 - Z)^{-4} \\
\mathcal{C}^{(0,2)}(Z) &= \frac{1}{8} (1 - Z)^{-4} \\
\mathcal{C}^{(0,3)}(Z) &= \frac{1}{4} (3-Z) (1 - Z)^{-5} \\
\mathcal{C}^{(1,1)}(Z) &= 12 (1 + 4 Z^2 - Z^4) (1 + Z)^{-4} (1 - Z)^{-4} \ln\left[ \frac{1}{2} (1 - Z) \right] \\
&\qquad+ (21 + 28 Z + 52 Z^2 - 28 Z^3 - 25 Z^4) (1 + Z)^{-3} (1 - Z)^{-5} \\
\mathcal{C}^{(1,2)}(Z) &= 12 Z (1 + Z^2) (1 + Z)^{-4} (1 - Z)^{-4} \ln\left[ \frac{1}{2} (1 - Z) \right] \\
&\qquad+ (8 + 35 Z + 28 Z^2 + 25 Z^3) (1 + Z)^{-3} (1 - Z)^{-5} \\
\mathcal{C}^{(1,3)}(Z) &= 12 Z (7 + 10 Z^2 - Z^4) (1 + Z)^{-5} (1 - Z)^{-5} \ln\left[ \frac{1}{2} (1 - Z) \right] \\
&\qquad+ (30 + 177 Z + 142 Z^2 + 184 Z^3 - 28 Z^4 - 25 Z^5) (1 + Z)^{-4} (1 - Z)^{-6} \eqend{.}
\end{split}
\end{equation}

Even though it seems that those coefficient functions are singular as $Z \to -1$ (which corresponds to antipodal points), this is only apparent. Expanding the logarithm around $Z = -1$ as
\begin{equation}
\label{ln_zone_expansion}
\ln\left[ \frac{1}{2}(1-Z) \right] = - \sum_{k=1}^\infty \frac{(1+Z)^k}{2^k \, k} \eqend{,}
\end{equation}
we get
\begin{equation}
\label{weyl_coeffs_finite}
\mathcal{C}^{(1,1)}(Z) = \frac{11}{128} + \bigo{Z+1} \eqend{,} \quad \mathcal{C}^{(1,2)}(Z) = \frac{23}{768} + \bigo{Z+1} \eqend{,} \quad \mathcal{C}^{(1,3)}(Z) = \frac{19}{240} + \bigo{Z+1} \eqend{,}
\end{equation}
which are perfectly regular at $Z = 1$. This means that the Weyl tensor correlator is only singular when the two points $x$ and $x'$ in the arguments of the correlator are connected by a null geodesic, with $Z(x,x') = 1$.

Of course, we can also express the basic bitensor set \eqref{basisbitensors_weyl} for the two-point Weyl correlation function in terms of the complete set $n_a$, $n_{a'}$ and $\hat{g}_{ab'}$ using equation~\eqref{relation_z_n}. One can then easily obtain
\begin{equation}
\begin{split}
{}^{(1)}\mathcal{C}_{abcdm'n'p'q'}(Z) &= - 2 Z \,{}^{(1)}\mathcal{D}_{abcdm'n'p'q'} + (1-Z)^2 \,{}^{(2)}\mathcal{D}_{abcdm'n'p'q'} \eqend{,} \\
{}^{(2)}\mathcal{C}_{abcdm'n'p'q'}(Z) &= - (3 + Z^2) \,{}^{(1)}\mathcal{D}_{abcdm'n'p'q'} - (3-Z)(1-Z) \,{}^{(2)}\mathcal{D}_{abcdm'n'p'q'} \\
&\qquad+ (1 - Z^2) \,{}^{(3)}\mathcal{D}_{abcdm'n'p'q'} \eqend{,} \\
{}^{(3)}\mathcal{C}_{abcdm'n'p'q'}(Z) &= (1 - Z^2) \,{}^{(1)}\mathcal{D}_{abcdm'n'p'q'} + (1-Z)^2 (1+Z) \,{}^{(2)}\mathcal{D}_{abcdm'n'p'q'} \eqend{,}
\end{split}
\end{equation}
where the (traceless) tensors ${}^{(k)}\mathcal{D}$ are defined as
\begin{equation}
\begin{split}
{}^{(1)}\mathcal{D}_{abcdm'n'p'q'} &= \left[ {}^{(1)}\mathcal{N} - 3 {}^{(5)}\mathcal{N} + {}^{(9)}\mathcal{N} \right]_{abcdm'n'p'q'} \\
{}^{(2)}\mathcal{D}_{abcdm'n'p'q'} &= \left[ - 12 {}^{(3)}\mathcal{N} - 3 {}^{(4)}\mathcal{N} + 12 {}^{(7)}\mathcal{N} + 2 {}^{(8)}\mathcal{N} \right]_{abcdm'n'p'q'} \\
{}^{(3)}\mathcal{D}_{abcdm'n'p'q'} &= \left[ - 2 {}^{(1)}\mathcal{N} + 6 {}^{(2)}\mathcal{N} - 12 {}^{(3)}\mathcal{N} + 3 {}^{(4)}\mathcal{N} + 3 {}^{(5)}\mathcal{N} - 6 {}^{(6)}\mathcal{N} + 12 {}^{(7)}\mathcal{N} \right]_{abcdm'n'p'q'} \eqend{,}
\end{split}
\end{equation}
with the basic bitensor set ${}^{(k)}\mathcal{N}$ introduced in equation~\eqref{basisbitensors_n}. The two-point Weyl correlation function \eqref{weyl_correlator} then reads
\begin{equation}
\label{weyl_correlator_d}
\begin{split}
\expect{{\tilde{C}^{ab}}{}_{cd}(x) {\tilde{C}^{m'n'}}{}_{p'q'}(x')} &= \frac{4 \kappa^2 H^6}{\mathpi^2} \sum_{k=1}^3 {}^{(k)}\mathcal{D}^{[ab]}{}_{[cd]}{}^{[m'n']}{}_{[p'q']} \distlim_{\epsilon \to 0^+} \bigg[ \frac{\alpha}{4} \kappa^2 H^2 \mathcal{D}^{(1,k)}(Z - \mathi\sgneps) \\
&+ \left( 1 + 6 \alpha \kappa^2 H^2 \left( \ln\left( \frac{\bar{\mu}}{H} \right) + \gamma \right) - \left( 5 \alpha - 2 \beta \right) \kappa^2 H^2 \right) \mathcal{D}^{(0,k)}(Z - \mathi\sgneps) \bigg] \eqend{,}
\end{split}
\end{equation}
where the coefficient functions $\mathcal{D}^{(0,k)}$ and $\mathcal{D}^{(1,k)}$ are given by
\begin{equation}
\label{weyl_n_basis_coeffs}
\begin{split}
\mathcal{D}^{(0,1)}(Z) &= \frac{1}{8} (3-Z) (1 - Z)^{-3} \\
\mathcal{D}^{(0,2)}(Z) &= \frac{1}{8} (7-Z) (1 - Z)^{-3} \\
\mathcal{D}^{(0,3)}(Z) &= \frac{1}{8} (1+Z) (1 - Z)^{-3} \\
\mathcal{D}^{(1,1)}(Z) &= 24 Z (1 + Z)^{-3} (1 - Z)^{-3} \ln\left[ \frac{1}{2} (1 - Z) \right] + 6 (1+5Z) (1 + Z)^{-2} (1 - Z)^{-4} \\
\mathcal{D}^{(1,2)}(Z) &= 12 (1+2Z+3Z^2) (1 + Z)^{-3} (1 - Z)^{-3} \ln\left[ \frac{1}{2} (1 - Z) \right] \\
&\qquad+ 3 (9+20Z+19Z^2) (1 + Z)^{-2} (1 - Z)^{-4} \\
\mathcal{D}^{(1,3)}(Z) &= 12 Z (1 + Z^2) (1 + Z)^{-3} (1 - Z)^{-3} \ln\left[ \frac{1}{2} (1 - Z) \right] \\
&\qquad+ (8 + 35 Z + 28 Z^2 + 25 Z^3) (1 + Z)^{-2} (1 - Z)^{-4} \eqend{.}
\end{split}
\end{equation}
The contracted Weyl correlator defined in equation~\eqref{quadrat_correlator_curvature} reduces in turn to
\begin{equation}
\label{quadrat_correlator_weyl}
\begin{split}
\expect{\mathrm{\tilde{W}eyl}{}^2(x,x')} &= - \frac{180 \alpha \kappa^4 H^8}{\mathpi^2} (1 + Z)^{-3} \ln\left[ \frac{1}{2} (1 - Z + \mathi \sgneps) \right] \\
&\quad- \frac{15 \alpha \kappa^4 H^8}{\mathpi^2} (7 - 35 Z + 29 Z^2 - 25 Z^3) (1 + Z)^{-2} (1 - Z + \mathi \sgneps)^{-4} \eqend{.}
\end{split}
\end{equation}
Again, the singularity for $Z \to -1$ is only apparent, as can be seen by expanding the logarithm around $Z=-1$ using equation~\eqref{ln_zone_expansion}.

For a general metric $g_{ab}$ the second Bianchi identity for the Riemann tensor (which also applies to its correlator) 
gives
\begin{equation}
\label{bianchi_weyl}
\nabla^a C_{abcd} = \nabla_{[c} \left[ R_{d]b} - \frac{1}{6} g_{d]b} R \right] \eqend{.}
\end{equation}
For the background de Sitter space $\hat{R}_{ab} = \Lambda \hat{g}_{ab}$ and $\hat{C}_{abcd} = 0$, so that both sides of  equation~\eqref{bianchi_weyl} vanish in that case and the equation is trivially satisfied.
Furthermore, without matter contributions $\hat{\nabla}_a \tilde{C}{}^{ab}{}_{cd} = 0$ and the Weyl correlator at tree level ($\alpha = \beta = 0$) must, therefore, fulfill this transversality condition as well. Setting $\alpha = \beta = 0$ in the explicit expression~\eqref{weyl_correlator} [or \eqref{weyl_correlator_d}], one can check that this is indeed the case. On the other hand, when the matter contributions are included, the Bianchi identity~\eqref{bianchi_weyl} relates the Weyl and Ricci-Weyl correlators at order $\kappa^4$:
\begin{equation}
\label{bianchi_weyl-weyl}
\hat{\nabla}_a \expect{\tilde{C}{}^{ab}{}_{cd}(x) \tilde{C}{}^{m'n'}{}_{p'q'}(x')} = \left( - \delta^b_s \delta_{[c}^t + \frac{1}{6} \delta_s^t \delta_{[c}^b \right) \hat{\nabla}_{d]} \expect{\tilde{R}{}^s{}_t(x) \tilde{C}{}^{m'n'}{}_{p'q'}(x')} \eqend{.}
\end{equation}
which we have explicitly checked to be the case for our result.

We now proceed to analyze the behavior of the Weyl correlator \eqref{weyl_correlator_d} in various interesting limits. First, for $Z \to -1$ we have already seen its regular behavior in \eqref{weyl_coeffs_finite}. From equation~\eqref{weyl_n_basis_coeffs} we see that in the light-cone limit $Z \to 1$, the components of the two-point Weyl correlator diverge like
\begin{equation}
\label{weyl_divergence}
\sim \kappa^2 H^6 (1 - Z + \mathi \sgneps)^{-3} + \kappa^4 H^8 (1 - Z + \mathi \sgneps)^{-4} + \kappa^4 H^8 (1 - Z + \mathi \sgneps)^{-3} \ln\left[ \frac{1}{2} (1 - Z + \mathi \sgneps) \right] \eqend{.}
\end{equation}
The quantum corrections due to the interaction with matter fields are, therefore, more singular near the light cone than the tree-level result, which corresponds to the first term in~\eqref{weyl_divergence}.
Note that throughout the rest of this section whenever writing the dominant terms for the large or the short-distance limits, indicated with the symbol $\sim$, we will leave out the dimensionless coefficients independent of any dimensionful parameters.

In order to analyze the behavior for \emph{large spacelike separations}, it is convenient to set the times $\eta$ and $\eta'$ of the two points $x$ and $x'$ equal by means of a de Sitter isometry, which is always possible for spacelike separated points.
The physical distance $d(x,x')$ between those points on the spatial section is then given by $d^2 = a^2(\eta) (\vec{x}-\vec{x}')^2$. From the explicit expression of $Z$ given in~\eqref{z_flat_coordinates}, this distance can be written as $d^2 = 2 H^{-2} (1-Z)$, which is equal to the squared Minkowski distance between those points in the embedding space, as seen from equation~\eqref{minkowski_z}. Hence, the limit $(\vec{x}-\vec{x}')^2 \to \infty$ for \emph{large spacelike separations} (i.e., superhorizon scales with $d \gg H^{-1}$) corresponds to $Z \to -\infty$, and we see that the components which decay more slowly go like
\begin{equation}
\label{weylweyl_decay}
\sim \kappa^2 H^6 \abs{Z}^{-2} + \kappa^2 H^6 \abs{Z}^{-3}
+ \kappa^4 H^8 \abs{Z}^{-2} + \kappa^4 H^8 \abs{Z}^{-3} \ln \abs{Z} + \ldots \eqend{,}
\end{equation}
which in terms of the proper physical distance $d$ reads
\begin{equation}
\label{weylweyl_decay2}
\sim \kappa^2 H^2 d^{-4} + \kappa^2 d^{-6} + \kappa^4 H^4 d^{-4} + \kappa^4 H^2 d^{-6} \ln (Hd)
+ \ldots \eqend{.}
\end{equation}
(Remember that $\kappa = \sqrt{16 \mathpi G_\text{N}} = \sqrt{16 \mathpi} \, l_\text{P}$, where $l_\text{P}$ is the Planck length.) It is worth pointing out that the second and last terms in \eqref{weylweyl_decay2} can be combined in a way which suggests the following relationship to a power law with a perturbative correction to the exponent:
\begin{equation}
\label{secular}
\kappa^2 d^{-6 + \kappa^2 H^2} \sim \kappa^2 d^{-6} \Big[ 1 +  \kappa^2 H^2 \ln (Hd)
+ \bigo{\kappa^2 H^2 \ln^2 (Hd)} \Big]  \eqend{.}
\end{equation}
This shows that even for $\kappa^2 H^2 \ll 1$ the perturbative expansion can break down for sufficiently large separations. Thus, in general one needs a nonperturbative calculation in $\kappa^2 H^2$ which resums the logarithms to capture correctly the behavior for large separations (and the corrections to it). The situation is analogous to the existence of secular terms in certain perturbative calculations when considering the evolution for a sufficiently long time \cite{fprv2013}.

For the \emph{short-distance} limit when the two points are spacelike separated, i.e., at subhorizon scales $d \ll H^{-1}$, one should recover the flat-space result. Indeed, using the previous expression connecting $d$ and $Z$ for points at equal times, one can see that $d \to 0$ implies $Z \to 1$, and that the logarithmic terms in~\eqref{weyl_n_basis_coeffs} are subdominant in this limit:
\begin{equation}
\kappa^4 H^8 (1 - Z)^{-3} \ln\left[ \frac{1}{2} (1 - Z) \right] \sim \kappa^4 H^2 d^{-6} \ln\left( \frac{1}{2} H^2 d^2 \right) \ll \kappa^4 d^{-8} \sim \kappa^4 H^8 (1 - Z)^{-4} \eqend{.}
\end{equation}
The dominant terms do not contain $H$, as one would expect, and we obtain
\begin{equation}
\label{weyl_shortdistance}
\sim \kappa^2 d^{-6} + \kappa^4 d^{-8} + \ldots \eqend{.}
\end{equation}
Instead, although the logarithmic term in \eqref{weyl_divergence} contributes to the divergent behavior in the limit $Z \to 1$, it does not contribute to the flat-space limit because it contains extra positive powers of $H$ after writing the result in terms of the distance $d$.
(Further details on the flat-space limit and the full expressions of the curvature tensor correlators in Minkowski space are provided in subsection~\ref{curvature_flat}.)
Furthermore, we see that at large distances (superhorizon scales) the Weyl correlator in de Sitter falls off, as given by \eqref{weylweyl_decay2}, more slowly than in flat space, characterized by \eqref{weyl_shortdistance}. This was also found for the stress tensor correlator for minimally coupled scalar fields (both massless and massive) \cite{perez-nadal10}.

For \emph{large timelike separations}, we can achieve $\vec{x} = \vec{x}'$ through a de Sitter isometry. The expression of $Z$ in spatially flat coordinates~\eqref{z_flat_coordinates} then shows that $\abs{\eta-\eta'} \to \infty$ corresponds to $Z \to \infty$, and we get the result~\eqref{weylweyl_decay}. The proper time elapsed along a geodesic connecting the two points is given by $\tau^2 = -\mu^2(x,x')$, where the relation between the geodesic distance $\mu$ and the biscalar $Z$ is given by equation~\eqref{relation_mu_z}. In the limit $Z \to \infty$ it follows that $\tau = H^{-1} \ln Z$, and therefore the components of the two-point Weyl correlation function which fall off more slowly go like
\begin{equation}
\sim \kappa^2 H^6 \mathe^{- 2 H \tau} + \kappa^4 H^8 \mathe^{- 2 H \tau} + \kappa^4 H^9 \tau \mathe^{- 3 H \tau} + \ldots \eqend{,}
\end{equation}
i.e., they fall off exponentially with the proper-time separation.

\subsection{The Ricci and Ricci-Weyl correlators}
\label{curvature_ricci}

The correlator of the linearized Ricci tensor as well as the Ricci-Weyl correlator can also be easily obtained from the Riemann correlator. Suitably contracting indices, we get for the two-point Ricci correlator
\begin{equation}
\label{ricci_correlator}
\begin{split}
\expect{{\tilde{R}^b}{}_d(x) {\tilde{R}^{n'}}{}_{q'}(x')} &= \frac{15 \alpha \kappa^4 H^4}{4 \mathpi^2} \bigg[ - H^4 \delta^b_d \delta^{n'}_{q'} (1-Z)^2 + 4 Z^{;b} Z_{;d} Z^{;n'} Z_{;q'} \\
&\qquad+ 2 \left( Z^{;b} Z_{;d}^{;n'} Z_{;q'} + Z_{;d} Z^{;bn'} Z_{;q'} + Z^{;b} Z_{;dq'} Z^{;n'} + Z_{;d} Z^{;b}_{;q'} Z^{;n'} \right) (1-Z) \\
&\qquad+ 2 \left( Z^{;bn'} Z_{;dq'} + Z^{;b}_{;q'} Z_{;d}^{;n'} \right) (1-Z)^2 \bigg] \distlim_{\epsilon \to 0^+} (1 - Z + \mathi \sgneps)^{-6} \eqend{.}
\end{split}
\end{equation}
On the other hand, using equation \eqref{weyl_projection_riemann}, for the Ricci-Weyl correlator we obtain
\begin{equation}
\label{ricciweyl_correlator}
\begin{split}
&\expect{{\tilde{R}^b}{}_d(x) {\tilde{C}^{m'n'}}{}_{p'q'}(x')} = \frac{3 \alpha \kappa^4 H^4}{\mathpi^2} \bigg[ 2 H^4 \delta^b_d \delta^{[m'}_{[p'} \delta^{n']}_{q']} (1-Z^2) - 2 H^2 Z^{;b} Z_{;d} \delta^{[m'}_{[p'} \delta^{n']}_{q']} \\
&\qquad- 6 H^2 \left( \delta^b_d - H^{-2} Z^{;b} Z_{;d} \right) \delta^{[m'}_{[p'} Z^{;n']} Z_{;q']} - 3 \delta^{[m'}_{[p'} \left( Z^{;b}_{;q']} Z_{;d}^{;n']} + Z^{;n']b} Z_{;q']d} \right) (1-Z^2) \\
&\qquad- 3 \delta^{[m'}_{[p'} \left( Z_{;q']} Z^{;n']b} Z_{;d} + Z^{;n']} Z^{;b}_{;q']} Z_{;d} + Z_{;q']} Z_{;d}^{;n']} Z^{;b} + Z^{;n']} Z_{;q']d} Z^{;b} \right) Z \\
&\qquad+ 6 H^{-2} Z^{;[m'} Z_{;[p'} \left( Z^{;b}_{;q']} Z_{;d}^{;n']} + Z^{;n']b} Z_{;q']d} \right) \bigg] \distlim_{\epsilon \to 0^+} (1 - Z + \mathi \sgneps)^{-5} \eqend{.}
\end{split}
\end{equation}
Given that $Z(x,x') = Z(x',x)$, the manifestly de Sitter-invariant results \eqref{riemann_correlator_pos} and \eqref{riemann_correlator_nn} for the Riemann correlator remain invariant if one swaps the two Riemann tensor operators. (In doing so, $\epsilon^\pm$ will change sign, which is relevant when $x$ and $x'$ are timelike separated, but not for spacelike separations: in that case both operators commute, in agreement with the microcausality principle.)
This implies that the result for the Weyl-Ricci correlator $\expect{{\tilde{C}^{mn}}{}_{pq}(x) {\tilde{R}^{b'}}{}_{d'}(x')}$ can be simply obtained from the right-hand side of \eqref{ricciweyl_correlator} by replacing primed indices with unprimed ones and vice versa.

It should be emphasized that there are no tree-level contributions to connected correlators involving the Ricci tensor. This can be understood by considering the Einstein equation for the Ricci operator regarded as a Heisenberg equation of motion. The cosmological constant source term is proportional to $\delta^a_b$ and its effect was entirely included in the background Ricci tensor, which was also the case for the expectation value of the stress tensor operator evaluated on the background metric. Hence, they do not contribute to the connected correlator, which only gets contributions starting at order $\kappa^2$ and involving stress tensor terms that give rise to matter loops when combined with similar terms from the other Ricci or Weyl operator (the terms that contribute at tree level to the Weyl-Weyl correlator cannot combine to form connected Feynman diagrams in this case). The first nonvanishing contributions to connected correlators involving at least one Ricci tensor are, therefore, of order $\kappa^4$ and one loop in the matter fields.

Note also that further contractions vanish, i.e., the two-point function of the Ricci scalar with any other curvature tensor is zero. This is a peculiarity of the conformal case which reflects the fact that the trace of the stress tensor (entirely due to the trace anomaly) does not fluctuate, and had already been found in Minkowski spacetime\cite{martin99}. For the contracted Ricci correlator defined in equation~\eqref{quadrat_correlator_curvature} we obtain
\begin{equation}
\label{quadrat_correlator_ricci}
\expect{\mathrm{\tilde{R}ic}{}^2(x,x')} = \frac{45 \alpha \kappa^4 H^8}{\mathpi^2} \distlim_{\epsilon \to 0^+} (1 - Z + \mathi \sgneps)^{-4} \eqend{.}
\end{equation}

In the bitensor basis furnished by the normal vectors $n_a$ and $n_{a'}$ and the parallel propagator $\hat{g}_{ab'}$, the above correlators \eqref{ricci_correlator} and \eqref{ricciweyl_correlator} read
\begin{equation}
\label{ricci_correlator_n}
\begin{split}
\expect{{\tilde{R}^b}{}_d(x) {\tilde{R}^{n'}}{}_{q'}(x')} &= \frac{15 \alpha \kappa^4 H^8}{4 \mathpi^2} \bigg[ - \delta^b_d \delta^{n'}_{q'} + 16 n^b n_d n^{n'} n_{q'} + 2 \left( \hat{g}^{bn'} \hat{g}_{dq'} + \hat{g}^b_{q'} \hat{g}_d^{n'} \right) \\
&\qquad+ 4 \left( n^b \hat{g}_d^{n'} n_{q'} + n_d \hat{g}^{bn'} n_{q'} + n^b \hat{g}_{dq'} n^{n'} + n_d \hat{g}^b_{q'} n^{n'} \right) \bigg] \times \\
&\qquad\qquad\times \distlim_{\epsilon \to 0^+} (1 - Z + \mathi \sgneps)^{-4}
\end{split}
\end{equation}
and
\begin{equation}
\label{ricciweyl_correlator_n}
\begin{split}
\expect{{\tilde{R}^b}{}_d(x) {\tilde{C}^{m'n'}}{}_{p'q'}(x')} &= \frac{3 \alpha \kappa^4 H^8}{\mathpi^2} \bigg[ 2 \left( \delta^b_d - n^b n_d \right) \delta^{[m'}_{[p'} \delta^{n']}_{q']} - 6 \delta^b_d \delta^{[m'}_{[p'} n^{n']} n_{q']} \\
&\qquad- 6 \delta^{[m'}_{[p'} \left( n_{q']} \hat{g}^{n']b} n_d + n^{n']} \hat{g}_{q']}^b n_d + n_{q']} \hat{g}_d^{n']} n^b + n^{n']} \hat{g}_{q']d} n^b \right) Z \\
&\qquad- 3 \left( \delta^{[m'}_{[p'} - 2 n^{[m'} n_{[p'} \right) \left( \hat{g}^b_{q']} \hat{g}_d^{n']} + \hat{g}^{n']b} \hat{g}_{q']d} \right) \bigg] \times \\
&\qquad\qquad\times (1+Z) \distlim_{\epsilon \to 0^+} (1 - Z + \mathi \sgneps)^{-4} \eqend{,}
\end{split}
\end{equation}
respectively. Both correlators are only singular as $Z \to 1$, where the most singular term goes like
\begin{equation}
\sim \kappa^4 H^8 (1 - Z + \mathi \sgneps)^{-4} \eqend{.}
\end{equation}
For \emph{large time-} and \emph{space-like separations}, the components which fall off more slowly go like
\begin{equation}
\sim \kappa^4 H^8 Z^{-4}
\end{equation}
for the two-point Ricci correlator and like
\begin{equation}
\sim \kappa^4 H^8 Z^{-2}
\end{equation}
for the Ricci-Weyl correlator. Similarly to the Weyl correlator case and taking into account that for large spacelike separations $\abs{Z} \sim H^2 d^2$, one can express this in terms of the proper physical distance $d$, while for large timelike separations the relation $Z \sim \exp(H \tau)$ allows to express the result in terms of the proper time $\tau$ along a geodesic connecting $x$ and $x'$.

\subsection{Result for general CFTs}
\label{general_CFT}
In this subsection we explain how the results for the Riemann correlator (and the related results for Weyl and Ricci correlators) obtained above for free scalar fields can be extended to general CFTs.
In order to do so, we need to go back to equations \eqref{seff_cft}-\eqref{effective_action_s2k} before particularizing the values of the constants $b$ and $b'$ to the case of free scalar fields.
We are then in a position to start analyzing how the different coefficients in expression \eqref{riemann_correlator_pos} will change in the general case.

We begin with some remarks about the different contributions to the Riemann correlator which will be useful for our reasoning below. The sum of terms involving the biscalars $\mathcal{R}^{(0,k)}$ in equation~\eqref{riemann_correlator_pos} corresponds to the tree-level result (of order $\kappa^2$) except for the global coefficient, which includes corrections of order $\kappa^4$.
Furthermore, since the terms proportional to the parameter $b'$ in the effective action $S_\text{eff}$ are either quadratic in $h^+_{ab}$ or quadratic in $h^-_{ab}$, they do not contribute to the component $V_{-+}$ of the matrix $V_{AB}$ defined in equation~\eqref{perturbation_v_definition}. From equation~\eqref{wightman_vector0} one can then conclude that there are no contributions from scalar and vectorial metric perturbations to the Riemann correlator which are proportional to $b'$. (Exactly the same argument applies to the terms proportional to $\ln \bar{\mu}$.)

Proceeding in the same way employed to derive equation~\eqref{tensor_S_G} for the tensorial perturbations, one obtains the following result for arbitrary values of $b$ and $b'$:
\begin{equation}
\label{tensor_S_G2}
\begin{split}
S_2^\text{G,TT}[h_{mn}] &= - \frac{1}{2} (2 b + b' - \beta) H^2 \int a^2 ( \partial_s h_{mn} ) ( \partial^s h^{mn} ) \total^4 x - b H^2 \int a^2 h'_{mn} h^{\prime mn} \total^4 x \\
&\quad+ \frac{b}{2} \int ( \dalembert h_{mn} ) ( \dalembert h^{mn} ) \ln a \total^4 x \eqend{.}
\end{split}
\end{equation}
When deriving this result, the terms proportional to $b'$ give rise to a shift of the cosmological constant term (at all orders in $h_{ab}$) which is included in the corrected cosmological constant $\Lambda_\text{eff}$ used to determine the background de Sitter metric as mentioned in footnote~\ref{Lambda_eff} and described in~\cite{frv2011a}.
On the other hand, the first integral on the right-hand side of equation~\eqref{tensor_S_G2} coincides with that in $S_0^\text{TT}$ and it will give a contribution to the Riemann correlator of the same form as the tree-level result. Hence, comparing with equation~\eqref{tensor_S_G}, one can see that the factor $(5 \alpha - 2 \beta)$ in equation \eqref{riemann_correlator_pos} should be replaced with $(4 b + 2 b' - 2 \beta)$. Since $b'$ does not appear anywhere else and does not contribute to the scalar and vectorial correlators, one can use the substitution $3 \alpha / 2 = b$ for all the remaining instances of $\alpha$, which also include the contributions from the nonlocal term $S_2^\text{K}[h^\pm]$, and the Riemann correlator for the general case becomes
\begin{equation}
\label{riemann_correlator_pos_general_CFT}
\begin{split}
&\expect{{\tilde{R}^{ab}}{}_{cd}(x) {\tilde{R}^{m'n'}}{}_{p'q'}(x')} = \\
&\quad\frac{4 \kappa^2 H^6}{\mathpi^2} \sum_{k=1}^9 {}^{(k)}\mathcal{R}^{[ab]}{}_{[cd]}{}^{[m'n']}{}_{[p'q']}(Z(x,x')) \distlim_{\epsilon \to 0^+} \bigg[ b \hspace{1pt} \kappa^2 H^2 \mathcal{R}^{(1,k)}(Z(x,x') - \mathi\sgneps) \\
&\qquad\quad+ \left( 1 + 4 \hspace{1pt} b \hspace{1pt} \kappa^2 H^2 \left( \ln\left( \frac{\bar{\mu}}{H} \right) + \gamma \right) - 2 \left( 2b + b' - \beta \right) \kappa^2 H^2 \right) \mathcal{R}^{(0,k)}(Z(x,x') - \mathi\sgneps) \bigg] \eqend{,}
\end{split}
\end{equation}
where the various objects are defined after equation \eqref{riemann_correlator_pos} and  for arbitrary values of the renormalization scale $\mu$ one needs to make the replacement $b \hspace{0.5pt} \ln\bar\mu \to (3/2)N a_1^\text{ren}(\mu) + b \hspace{0.5pt} \ln \mu$, as explained in section~\ref{2pf_action}.

Using the same kind of substitutions, one gets analogous results for equation \eqref{riemann_correlator_nn} and for expressions \eqref{weyl_correlator} and \eqref{weyl_correlator_d} for the Weyl correlator. Moreover, since the Ricci and Ricci-Weyl correlators, obtained by suitable contractions of the Riemann correlator, vanish for the tree-level result, they are entirely determined by the terms involving the biscalars $\mathcal{R}^{(1,k)}$. Therefore, one simply needs to make the replacement $3 \alpha / 2 = b$ in the results for the Ricci and Ricci-Weyl correlators of section \ref{curvature_ricci}.

\subsection{The flat-space limit}
\label{curvature_flat}

For points separated by distances much smaller than the Hubble length $H^{-1}$ one expects that the two-point correlation function of the Riemann tensor should reproduce the Minkowski results. Thus, taking the curvature radius of de Sitter space $H^{-1}$ to infinity, i.e., the Hubble constant $H$ to zero, we can obtain the two-point correlation functions of the curvature tensors in a Minkowski background. In the Poincaré patch of de Sitter spacetime, this can be done by choosing the spatially flat coordinate system
\begin{equation}
\total s^2 = - \total t^2 + \mathe^{2 H t} \total \vec{x}^2 \eqend{,}
\end{equation}
expressing everything in terms of those coordinates and taking $H \to 0$ so that the metric goes over to the Minkowski metric. However, the flat space limit can also be obtained easily in a way independent of the choice of a particular coordinate system by noting that the geodesic distance $\mu(x,x')$ between two points $x$ and $x'$ in the Minkowski spacetime is simply given by $\mu(x,x') = \sqrt{(x-x')^2}$. For the invariant biscalar $Z$ given in equation~\eqref{relation_mu_z} we then calculate
\begin{equation}
Z(x,x') = 1 - \frac{1}{2} H^2 (x-x')^2 + \bigo{H^3} \eqend{.}
\end{equation}
For the covariant derivatives of $Z$ we have
\begin{equation}
\nabla_a Z \to - H^2 \eta_{ab} (x-x')^b + \bigo{H^3} \eqend{,} \qquad \nabla_a \nabla_{b'} Z \to H^2 \eta_{ab'} + \bigo{H^3} \eqend{,}
\end{equation}
and for the bitensor set of Allen and Jacobson we obtain from equation~\eqref{relation_z_n}
\begin{equation}
n_a \to \frac{\eta_{ab} (x-x')^b}{\sqrt{(x-x')^2}} + \bigo{H} \eqend{,} \qquad \hat{g}_{ab'} \to \eta_{ab'} + \bigo{H} \eqend{.}
\end{equation}

Thus, the Weyl correlator~\eqref{weyl_correlator} is in the flat-space limit given by
\begin{equation}
\label{weyl_correlator_flat}
\begin{split}
\expect{{\tilde{C}^{ab}}{}_{cd}(x) {\tilde{C}^{m'n'}}{}_{p'q'}(x')} &= \frac{8 \kappa^2}{\mathpi^2} \left[ {}^{(1)}\mathcal{W} + 3 {}^{(2)}\mathcal{W} + {}^{(3)}\mathcal{W} \right]{}^{[ab]}{}_{[cd]}{}^{[m'n']}{}_{[p'q']} \times \\
&\qquad\qquad\times \distlim_{\epsilon \to 0^+} ((x-x')^2 + \mathi\sgneps)^{-3} \\
&+ \frac{48 \alpha \kappa^4}{\mathpi^2} \left[ 3 {}^{(1)}\mathcal{W} + 12 {}^{(2)}\mathcal{W} + 8 {}^{(3)}\mathcal{W} \right]{}^{[ab]}{}_{[cd]}{}^{[m'n']}{}_{[p'q']} \times \\
&\qquad\qquad\times \distlim_{\epsilon \to 0^+} ((x-x')^2 + \mathi\sgneps)^{-4} \eqend{,}
\end{split}
\end{equation}
where the flat-space basis bitensors read
\begin{equation}
\begin{split}
{}^{(1)}\mathcal{W}_{abcdm'n'p'q'} &= \eta_{ac} \eta_{bd} \eta_{m'p'} \eta_{n'q'} - 6 \eta_{ac} \eta_{m'p'} \eta_{b(n'} \eta_{q')d} + 4 \eta_{a(m'} \eta_{p')c} \eta_{b(n'} \eta_{q')d} \\
{}^{(2)}\mathcal{W}_{abcdm'n'p'q'} &= - 12 \eta_{ac} \eta_{m'p'} y_b y_d y_{n'} y_{q'} - 12 \eta_{ac} \eta_{m'p'} y_{(b} \eta_{d)(n'} y_{q')} \\
&\qquad+ 24 y_a y_c y_{m'} y_{p'} \eta_{b(n'} \eta_{q')d} + 16 y_{(a} \eta_{c)(m'} y_{p')} \eta_{b(n'} \eta_{q')d} \\
{}^{(3)}\mathcal{W}_{abcdm'n'p'q'} &= - 2 \eta_{ac} \eta_{bd} \eta_{m'p'} \eta_{n'q'} + 6 \eta_{ac} \eta_{m'p'} \left( \eta_{bd} y_{n'} y_{q'} + y_b y_d \eta_{n'q'} \right) \\
&\qquad- 12 \eta_{ac} \eta_{m'p'} y_b y_d y_{n'} y_{q'} + 12 \eta_{ac} \eta_{m'p'} y_{(b} \eta_{d)(n'} y_{q')} + 6 \eta_{ac} \eta_{m'p'} \eta_{b(n'} \eta_{q')d} \\
&\qquad- 12 \left( \eta_{ac} y_{m'} y_{p'} + y_a y_c \eta_{m'p'} \right) \eta_{b(n'} \eta_{q')d} + 24 y_a y_c y_{m'} y_{p'} \eta_{b(n'} \eta_{q')d} \eqend{.}
\end{split}
\end{equation}
with unit vectors $y_a$ defined as $y_a = \eta_{ab} (x-x')^b ((x-x')^2)^{-1/2}$.
Note that the coefficient $\beta$ as well as the renormalization scale $\mu$ do not appear, i.e., the flat-space limit (for conformally coupled scalar fields) is insensitive to the contributions from counterterms quadratic in the curvature tensors.
To compare with the de Sitter results, note that for points $x$ and $x'$ at equal time $x^0 = (x')^0$, the physical distance is $d^2 = (\vec{x}-\vec{x}')^2 = (x-x')^2$, and the two-point Weyl correlation function falls off like
\begin{equation}
\sim \kappa^2 d^{-6} + \kappa^4 d^{-8} \eqend{,}
\end{equation}
which agrees with the short-distance limit for spacelike separations given by~\eqref{weyl_shortdistance}.

For the Ricci correlator~\eqref{ricci_correlator} we obtain
\begin{equation}
\label{ricci_correlator_flat}
\begin{split}
\expect{{\tilde{R}^b}{}_d(x) {\tilde{R}^{n'}}{}_{q'}(x')} &= \frac{60 \alpha \kappa^4}{\mathpi^2} \bigg[ - \delta^b_d \delta^{n'}_{q'} + 16 y^b y_d y^{n'} y_{q'} + 2 \left( \eta^{bn'} \eta_{dq'} + \eta^b_{q'} \eta_d^{n'} \right) \\
&\quad+ 4 \left( y^b \eta_d^{n'} y_{q'} + y_d \eta^{bn'} y_{q'} + y^b \eta_{dq'} y^{n'} + y_d \eta^b_{q'} y^{n'} \right) \bigg] \times \\
&\qquad\qquad\times \distlim_{\epsilon \to 0^+} ((x-x')^2 + \mathi \sgneps)^{-4} \eqend{,}
\end{split}
\end{equation}
which is of order $\kappa^4$ and falls off like $\kappa^4 d^{-8}$.
Similarly, the Ricci-Weyl correlation function~\eqref{ricciweyl_correlator} reads in the flat-space limit
\begin{equation}
\label{ricciweyl_correlator_flat}
\begin{split}
\expect{{\tilde{R}^b}{}_d(x) {\tilde{C}^{m'n'}}{}_{p'q'}(x')} &= \frac{96 \alpha \kappa^4}{\mathpi^2} \bigg[ 2 \left( \delta^b_d - y^b y_d \right) \delta^{[m'}_{[p'} \delta^{n']}_{q']} - 6 \delta^b_d \delta^{[m'}_{[p'} y^{n']} y_{q']} \\
&\qquad- 6 \delta^{[m'}_{[p'} \left( y_{q']} \eta^{n']b} y_d + y^{n']} \eta_{q']}^b y_d + y_{q']} \eta_d^{n']} y^b + y^{n']} \eta_{q']d} y^b \right) \\
&\qquad- 3 \left( \delta^{[m'}_{[p'} - 2 y^{[m'} y_{[p'} \right) \left( \eta^b_{q']} \eta_d^{n']} + \eta^{n']b} \eta_{q']d} \right) \bigg] \times \\
&\qquad\qquad\times \distlim_{\epsilon \to 0^+} ((x-x')^2 + \mathi \sgneps)^{-4} \eqend{.}
\end{split}
\end{equation}
which is also of order $\kappa^4$ and falls off like $\kappa^4 d^{-8}$.

As before, the Riemann correlator can be calculated from the decomposition~\eqref{riemann_correlator_decomp}. This result is manifestly Lorentz invariant and given in spacetime coordinates. To our knowledge, it has not appeared in the literature before, although the two-point correlator for the metric perturbations has been calculated in~\cite{martin99} in Fourier space.
Finally, the results for a general CFT can be easily obtained with the substitution $3 \alpha / 2 = b$ in all the results of this subsection, as discussed in section \ref{general_CFT}.

\section{Comparison with previous results}
\label{comparison}

Our result on the full Riemann correlator at order $\kappa^4$ (one-loop order for free matter fields) encompasses some previously calculated partial results. In particular, the Weyl correlator at tree level was calculated by Kouris in \cite{kouris}, and the stress tensor correlator, which is related through the Einstein equation to the Ricci correlator, was obtained by Osborn and Shore in \cite{osbornshore}. These are non-trivial checks of our results.

After correcting Kouris's result by means of suitable symmetrization, we find agreement up to a global factor of 2, while we agree completely with Osborn and Shore.

\subsection{Weyl-Weyl correlator at tree level}
\label{comparison_weyl}

In his calculation of the Weyl correlator at tree level
Kouris started with a de Sitter-invariant two-point function for the metric perturbations in closed coordinates \cite{higuchikouris}. His final result is given by
\begin{equation}
\expect{{\tilde{C}}_{abcd}(x) {\tilde{C}}_{m'n'p'q'}(x')} = \sum_{k=1}^7 D^{(k)}(z(x,x')) S^{(k)}_{[ab][cd][m'n'][p'q']} \eqend{,}
\end{equation}
where
\begin{equation}
\begin{aligned}
D^{(1)} &= \frac{4 G_\text{N} H^6}{\mathpi} \frac{12}{(z-1)^3} &=&\, - \frac{\kappa^2 H^6}{4 \mathpi^2} \frac{96}{(1 - Z)^3} \\
D^{(2)} &= \frac{4 G_\text{N} H^6}{\mathpi} \left( \frac{18}{(z-1)^3} - \frac{6}{(z-1)^2} \right) &=&\, - \frac{\kappa^2 H^6}{4 \mathpi^2} \left( \frac{144}{(1 - Z)^3} + \frac{24}{(1 - Z)^2} \right) \\
D^{(3)} &= \frac{4 G_\text{N} H^6}{\mathpi} \left( - \frac{6}{(z-1)^3} + \frac{6}{(z-1)^2} \right) &=&\, \frac{\kappa^2 H^6}{4 \mathpi^2} \left( \frac{48}{(1 - Z)^3} + \frac{24}{(1 - Z)^2} \right) \\
D^{(4)} &= \frac{4 G_\text{N} H^6}{\mathpi} \left( \frac{3}{(z-1)^3} + \frac{3}{(z-1)^3} \right) &=&\, \frac{\kappa^2 H^6}{4 \mathpi^2} \left( - \frac{24}{(1 - Z)^3} + \frac{12}{(1 - Z)^2} \right) \\
D^{(5)} &= \frac{4 G_\text{N} H^6}{\mathpi} \left( - \frac{3}{2 (z-1)^3} + \frac{3}{2 (z-1)^2} \right) &=&\, \frac{\kappa^2 H^6}{4 \mathpi^2} \left( \frac{12}{(1 - Z)^3} + \frac{6}{(1 - Z)^2} \right) \\
D^{(6)} &= \frac{4 G_\text{N} H^6}{\mathpi} \frac{3}{(z-1)^2} &=&\, \frac{\kappa^2 H^6}{4 \mathpi^2} \frac{12}{(1 - Z)^2} \\
D^{(7)} &= \frac{4 G_\text{N} H^6}{\mathpi} \left( \frac{1}{4 (z-1)^3} + \frac{3}{4 (z-1)^2} \right) &=&\, \frac{\kappa^2 H^6}{4 \mathpi^2} \left( - \frac{2}{(1 - Z)^3} + \frac{3}{(1 - Z)^2} \right) \\
\end{aligned}
\end{equation}
and
\begin{equation}
\begin{split}
S^{(1)}_{abcdm'n'p'q'} &= n_a n_c n_{m'} n_{p'} \left( \hat{g}_{bd} \hat{g}_{n'q'} - 2 \hat{g}_{bn'} \hat{g}_{dq'} \right) \\
S^{(2)}_{abcdm'n'p'q'} &= n_a n_{p'} \hat{g}_{bn'} \hat{g}_{cq'} \hat{g}_{dm'} \\
S^{(3)}_{abcdm'n'p'q'} &= n_c n_{p'} \hat{g}_{m'q'} \hat{g}_{bd} \hat{g}_{an'} \\
S^{(4)}_{abcdm'n'p'q'} &= n_{m'} n_{p'} \hat{g}_{bq'} \hat{g}_{ac} \hat{g}_{dn'} + n_a n_c \hat{g}_{dq'} \hat{g}_{bn'} \hat{g}_{m'p'} - \frac{1}{2} \hat{g}_{bd} \left( n_{m'} n_{p'} \hat{g}_{ac} \hat{g}_{n'q'} + n_a n_c \hat{g}_{n'q'} \hat{g}_{m'p'} \right) \\
S^{(5)}_{abcdm'n'p'q'} &= \hat{g}_{an'} \hat{g}_{bp'} \hat{g}_{cq'} \hat{g}_{dm'} \\
S^{(6)}_{abcdm'n'p'q'} &= \hat{g}_{ac} \hat{g}_{dm'} \hat{g}_{n'q'} \hat{g}_{bp'} \\
S^{(7)}_{abcdm'n'p'q'} &= \hat{g}_{ac} \hat{g}_{bd} \hat{g}_{m'p'} \hat{g}_{n'q'} \eqend{.}
\end{split}
\end{equation}
In writing the expressions for the coefficients $D^{(k)}$, we have replaced Kouris' $z$ with our $Z$ according to
\begin{equation}
z = \cos^2\left( \frac{1}{2} H \mu \right) = \frac{1}{2} \left( 1 + \cos(H \mu) \right) = \frac{1}{2} \left( 1 + Z \right) \eqend{.}
\end{equation}

In order to compare with our results, we can directly take the expression of the Weyl correlator  expressed in terms of normal vectors and parallel propagators in equation~\eqref{weyl_correlator_d}. However, we first note that Kouris' expression respects \emph{neither} the cyclic symmetry $C_{abcd} + C_{acdb} + C_{adbc} = 0$ \emph{nor} the symmetry under exchange of index pairs $C_{abcd} = C_{cdab}$. This was corrected in a later erratum~\cite{kouris2012}, and the correction is equivalent to replacing the Weyl tensor with the expression
\begin{equation}
\label{riemannization}
\begin{split}
C_{abcd} \to \frac{1}{12} &\big( C_{abcd} + C_{adcb} + C_{cbad} + C_{cdab} - C_{abdc} - C_{acdb} - C_{dbac} - C_{dcab} \\
&\quad- C_{bacd} - C_{bdca} - C_{cabd} - C_{cdba} + C_{badc} + C_{bcda} + C_{dabc} + C_{dcba} \big) \eqend{,}
\end{split}
\end{equation}
which can be shown to equal the Weyl tensor and manifestly respects all its symmetries~\cite{riemannsymmetry}. We then get
\begin{equation}
\begin{split}
S^{(1)}_{abcdm'n'p'q'} &= {}^{(3)}\mathcal{N}_{abcdm'n'p'q'} - {}^{(7)}\mathcal{N}_{abcdm'n'p'q'} \\
S^{(2)}_{abcdm'n'p'q'} &= - \frac{1}{12} {}^{(8)}\mathcal{N}_{abcdm'n'p'q'} \\
S^{(3)}_{abcdm'n'p'q'} &= - \frac{1}{4} {}^{(4)}\mathcal{N}_{abcdm'n'p'q'} \\
S^{(4)}_{abcdm'n'p'q'} &= - \frac{1}{2} {}^{(2)}\mathcal{N}_{abcdm'n'p'q'} + \frac{1}{2} {}^{(6)}\mathcal{N}_{abcdm'n'p'q'} \\
S^{(5)}_{abcdm'n'p'q'} &= \frac{1}{6} {}^{(9)}\mathcal{N}_{abcdm'n'p'q'} \\
S^{(6)}_{abcdm'n'p'q'} &= \frac{1}{2} {}^{(5)}\mathcal{N}_{abcdm'n'p'q'} \\
S^{(7)}_{abcdm'n'p'q'} &= {}^{(1)}\mathcal{N}_{abcdm'n'p'q'} \eqend{.}
\end{split}
\end{equation}
Additionally, the sign of $S^{(6)}_{abcdm'n'p'q'}$ as given by Kouris must be changed. Kouris' expression then reduces to
\begin{equation}
\begin{split}
\expect{{\tilde{C}}_{abcd}(x) {\tilde{C}}_{m'n'p'q'}(x')} &= \frac{2 \kappa^2 H^6}{4 \mathpi^2} \bigg[ \frac{1}{2} \frac{1-3Z}{(1 - Z)^3} {}^{(1)}\mathcal{N} + 3 \frac{1+Z}{(1 - Z)^3} {}^{(2)}\mathcal{N} - \frac{48}{(1 - Z)^3} {}^{(3)}\mathcal{N} \\
&\quad- 3 \frac{3-Z}{(1 - Z)^3} {}^{(4)}\mathcal{N} - 3 \frac{1}{(1 - Z)^2} {}^{(5)}\mathcal{N} - 3 \frac{1+Z}{(1 - Z)^3} {}^{(6)}\mathcal{N} \\
&\quad+ \frac{48}{(1 - Z)^3} {}^{(7)}\mathcal{N} + \frac{7-Z}{(1 - Z)^3} {}^{(8)}\mathcal{N} + \frac{1}{2} \frac{3-Z}{(1 - Z)^3} {}^{(9)}\mathcal{N} \bigg]_{[ab][cd][m'n'][p'q']} \eqend{,}
\end{split}
\end{equation}
which coincides with the tree-level part of our result \eqref{weyl_correlator_d} up to a factor of 2 if we interpret $1-Z$ as $1-Z+\mathi \sgneps$. Both the sign and the factor of 2 are acknowledged in the erratum~\cite{kouris2012}.

We thus see that the Bunch-Davies vacua for gravitons in closed and spatially flat coordinates are physically equivalent (at least for geometric properties within any region of finite proper size) since they lead to the same Weyl correlator -- despite the existence of an infrared divergence for the graviton propagator in flat coordinates~\cite{allen87} which is absent in closed coordinates~\cite{higuchikouris}. Indeed, the local geometric properties in a region of finite size are entirely characterized by the Riemann tensor in that region. Moreover, since the linear perturbation of the Ricci tensor with an appropriately raised index is gauge invariant and vanishes around a de Sitter background, the Weyl correlator is sufficient to characterize the fluctuations of the local geometry at this order. Only global geometric properties, which require knowledge over an infinite size region and to which the ``no-hair'' properties and attractor character of de Sitter space do not apply, are not properly characterized by the Riemann tensor.

Using a de Sitter-noninvariant gauge in conformally flat coordinates, Mora, Tsamis and Woodard have also calculated the two-point Weyl correlation function at tree level \cite{morawoodard12a,morawoodard12b}. In their latter article~\cite{morawoodard12b} they state an agreement with Kouris' result after the above corrections are taken into account. Our result, therefore, agrees also with theirs.

\subsection{The Ricci and stress tensor correlators}
\label{comparison_ricci}

We can compare our Ricci correlator to the stress tensor correlator for conformal fields obtained by Osborn and Shore~\cite{osbornshore}.
They use as a basic set of bitensors the geodesic distance $\mu$, the normal vectors $n_a$ and $n_{a'}$ and the parallel propagator $\hat{g}_{ab'}$, which they name $\theta$ (rescaled by $H$), $\hat{x}_a, \hat{y}_a$ and $I_{ab}$, respectively. Adapted to our notation, their result for the four-dimensional case is given by
\begin{equation}
\expect{T_{ab} T_{c'd'}} = \frac{1}{48 \mathpi^4} H^8 (1 - Z)^{-4} \left[ \left( \hat{g}_{a(c'} + 2 n_a n_{(c'} \right) \left( \hat{g}_{d')b} + 2 n_{d')} n_b \right) - \frac{1}{4} \hat{g}_{ab} \hat{g}_{c'd'} \right] \eqend{.}
\end{equation}
Since all connected correlators of the Ricci scalar vanish, as remarked in section~\ref{curvature_ricci}, up to the order at which we are working the Einstein equation gives
\begin{equation}
\hat{g}_{as} \hat{g}_{c'm'} \expect{\tilde{R}^s{}_b \tilde{R}^{m'}{}_{d'}} = \frac{1}{4} \kappa^4 \expect{T_{ab} T_{c'd'}} \eqend{,}
\end{equation}
which coincides exactly with equation \eqref{ricci_correlator_n} when we take $\alpha = (2880 \mathpi^2)^{-1}$, corresponding to the single scalar field that Osborn and Shore considered. Once again, we should interpret their $1-Z$ as $1-Z+\mathi \sgneps$.
Furthermore, taking into account the substitution $3 \alpha / 2 = b$ for the prefactor, as explained in section \ref{general_CFT}, one can see that the result also agrees for the other examples of CFTs that they studied.

\section{Discussion}
\label{discussion}

In this article, we have computed the full two-point Riemann correlator for metric perturbations around de Sitter spacetime including radiative corrections from loops of conformal matter fields and shown that they do not break de Sitter invariance. Specifically, we have obtained the exact result up to order $H^4/m_\text{p}^4$ and written it in a manifestly de Sitter-invariant form in terms of the invariant biscalar $Z(x,x')$ and its covariant derivatives (or, equivalently, in terms of the basis of invariant bitensors introduced by Allen and Jacobson~\cite{allenjacobson86}). We have primarily focused on free massless, conformally coupled scalar fields as a definite example. Nevertheless, our approach makes use of the CTP effective action that results from functionally integrating the matter fields and, as explained in section~\ref{2pf_action}, this effective action has the same form for any CFT up to two constant coefficients ($b$ and $b'$) which depend on the specific theory. Therefore, we were also able to obtain the general result valid for any CFT: for free theories it corresponds to one-loop order in the matter fields, whereas for interacting ones it accounts for the effects of processes involving any number of matter loops but no internal graviton propagators.

It should be emphasized that the de Sitter-invariant result for the Riemann correlator, which fulfills both Bianchi identities and all the relevant symmetries under exchange of indices, comes out at the end of a rather nontrivial and lengthy calculation where de Sitter invariance was not assumed at any point. In fact, in order to compute the correlator for the metric perturbations, we completely fixed the gauge using a non-invariant gauge fixing. (Hence, our calculation is not affected by the recent controversy concerning the use of average gauges in de Sitter~\cite{woodard2009,moratsamiswoodard2012,higuchi2011,faizal2012,higuchi2012,morrison2013}.)
However, what is always necessary so as to have an exactly de Sitter-invariant result is a suitable de Sitter-invariant state. In this respect it is important to keep in mind that while the Bunch-Davies vacuum for a free theory is de Sitter invariant, it will cease to be de Sitter invariant when it is evolved by a Hamiltonian that includes interactions (a point which has often been overlooked in the literature). A de Sitter-invariant state for the interacting theory is needed instead. The $\mathi \epsilon$ prescription introduced in section~\ref{2pf} selects as asymptotic initial state an adiabatic vacuum of the interacting theory which constitutes the appropriate generalization of the Bunch-Davies vacuum to the interacting case. 
In~\cite{higuchimarolf10} this kind of prescription for the expanding Poincar\'e patch was shown to be equivalent to the Hartle-Hawking state (also known as Euclidean vacuum) for the interacting theory, obtained by analytic continuation of the correlation functions calculated in the hypersphere to global de Sitter spacetime, a procedure which, if well defined, automatically leads to de Sitter-invariant results. This was done for massive scalar fields with nonderivative interactions. In contrast, the case studied here involves derivative interactions, graviton fields that behave like massless minimally coupled scalar fields and the gauge freedom associated with local diffeomorphisms. In this sense, one can regard our de Sitter-invariant result as supporting evidence that the Hartle-Hawking construction is also possible in this case and is equivalent to the calculation in the Poincar\'e patch (with some qualifications).

Besides the fact that it exhibits de Sitter invariance (or its absence) manifestly when written in terms of maximally symmetric bitensors, the Riemann correlator with appropriately raised indices has other appealing properties at the order at which we are working. It is gauge invariant at order $1/N$ for a large $N$ expansion in the number of matter fields (which excludes the contributions of graviton loops, of higher order in $1/N$). Furthermore, the Riemann tensor provides a suitable characterization of the local geometry and its correlator is an infrared-safe observable \cite{gerstenlauer11,giddingssloth2011,urakawa10,tanakaurakawa2013} at that order. In contrast, there are gauge-invariant quantities which cannot be determined by the geometry within a region of fixed (and finite) proper size: they depend on global properties or on geometric properties associated with a region whose proper size grows with the cosmological expansion (e.g., a region of fixed comoving size). Classical no-hair theorems for de Sitter space do not apply to them and in general the correlators characterizing their quantum fluctuations exhibit infrared divergences as the number of past e-foldings tends to infinity, which is incompatible with de Sitter invariance. (A simple example of such gauge-invariant objects is the tree-level correlator for the transverse-traceless metric perturbations in terms of the background spacetime coordinates.)
Thus, in the gravitational case de Sitter invariance can only hold at the quantum mechanical level for observables depending on the geometry within regions of arbitrary but finite proper size. The Riemann tensor and its correlators constitute a convenient example of such class of observables at leading order in $1/N$.

We close this section with an outlook of possible lines of future research.
First of all, there are several aspects concerning the case of conformal matter fields that one can further investigate. One of them is getting a nonperturbative result in $\kappa^2$ for the Riemann correlator. This is necessary in order to capture correctly the full behavior of the correlator at large separations. Indeed, as shown in equation~\eqref{secular}, one expects to get terms with arbitrarily high powers of $\kappa^2 H^2 \ln Z$, which cannot be treated as a small expansion parameter for sufficiently large values of $|Z|$, and one needs to resum all these logarithmic terms. Such a resummation would be automatically taken care of if one were able to obtain the nonperturbative result in $\kappa^2$ for the correlator, but still to order $1/N$ so that contributions from graviton loops do not need to be considered and the Riemann correlator is gauge invariant (and infrared safe).
Interestingly, it has been argued that the correlator of the metric perturbations to leading order in $1/N$ can be obtained in terms of the homogeneous solutions and retarded propagator of the linearized semiclassical Einstein equation \cite{hu04}. Combining that insight with the nonperturbative solutions, recently obtained in~\cite{fprv2013}, of the linearized semiclassical Einstein equation around a de Sitter background (which takes into account the backreaction of conformal fields) may make it possible to obtain a nonperturbative result in $\kappa^2$ for the Riemann correlator.

On the other hand, it is also of great interest to extend the results presented here to include the effects of radiative corrections from other kinds of fields besides conformal matter fields. A first step would be to consider loop corrections to the Riemann correlator due to free scalar fields with general mass and curvature-coupling parameter $\xi$. Massless (or sufficiently light) minimally coupled fields are particularly interesting since they play an important role in inflationary cosmology and are the ones leading to stronger infrared effects in de Sitter.
More challenging is to include the effects of graviton loops. The necessary renormalization for dealing with the UV divergences that arise in this case when functionally integrating the metric perturbations should be performed within a covariant treatment which requires in turn the use of DeWitt-Fadeev-Popov ghosts and BRST quantization~\cite{weinberg_v2}, and is technically quite involved.
Furthermore, once graviton loops are taken into account, the Riemann correlator (expressed in terms of background spacetime coordinates) ceases to be gauge invariant and infrared safe. In fact, when going beyond a linear treatment of the metric perturbations, one needs in general to confront the (conceptual) difficulties of constructing diffeomorphism invariant observables, which can be, at best, approximately local~\cite{giddings06}.
Nevertheless, for a perturbative calculation around the maximally symmetric de Sitter background, it may be possible to consider a generalized gauge-invariant Riemann correlator by considering the parallel transport (with the perturbed metric) of the two Riemann tensors to a common point. At sufficiently higher order, however, new UV divergences will arise due to the fact that the tensors are parallelly transported along a one-dimensional geodesic and some kind of smearing around the curve should be introduced.

\acknowledgments

M.~F.\ acknowledges financial support through FPU scholarship No.~AP2010-5453, as well as enlightening discussions with G. Pérez-Nadal. E.~V.\ and M.~F.\ also acknowledge partial financial support by the Research Projects MCI FPA2007-66665-C02-02, FPA2010-20807-C02-02, CPAN CSD2007-00042, with\-in the program Consolider-Ingenio 2010, and AGAUR 2009-SGR-00168.
A.~R.\ is supported by the Deutsches Zentrum f\"ur Luft- und Raumfahrt (DLR) with funds provided by the Bundesministerium f\"ur Wirtschaft und Technologie (BMWi) under Grant No.~50WM1136 (QUANTUS III).
For the tedious tensor algebra we used the open source \emph{xAct} tensor package~\cite{xact}.

\appendix

\section{Metric expansion}
\label{appendix_metric}

For the perturbed metric $g_{ab}$ we have
\begin{equation}
\begin{split}
g_{ab} &= \eta_{ab} + \kappa h_{ab} \\
g^{ab} &= \eta^{ab} - \kappa h^{ab} + \kappa^2 h^a_m h^{bm} + \bigo{h^3} \\
h &= \eta^{ab} h_{ab} \\
\sqrt{-g} &= 1 + \frac{1}{2} \kappa h + \frac{1}{8} \kappa^2 h^2 - \frac{1}{4} \kappa^2 h_{mn} h^{mn} + \bigo{h^3} \eqend{.}
\end{split}
\end{equation}
where all indices are raised and lowered with the unperturbed metric $\eta_{ab}$
and we have rescaled the metric perturbations $h_{ab}$ by a factor $\kappa = \sqrt{16 \mathpi G_\text{N}}$. Note that then the kinetic term is canonically normalized up to a factor $1/2$.

For the calculation of the Christoffel symbols (the connection) and the curvature tensors we regard $h_{ab}$ as a tensor field in flat space. The covariant derivative associated with $g_{ab}$ is denoted by $\nabla_c$. For the Christoffel symbols we obtain
\begin{equation}
\label{christoffel}
\begin{split}
\christoffel{a}{b}{c} &= \frac{1}{2} \kappa {S^a}_{bc} - \frac{1}{2} \kappa^2 h^a_m {S^m}_{bc} + \bigo{h^3} \\
{S^a}_{bc} &= \partial_b h^a_c + \partial_c h^a_b - \partial^a h_{bc} \\
{S^a}_{ac} &= \partial_c h \\
\partial_c h_{mb} &= \frac{1}{2} \eta_{an} \left( \delta^n_m {S^a}_{bc} + \delta^n_b {S^a}_{mc} \right)  \eqend{,}
\end{split}
\end{equation}
where $\partial^m = \eta^{mn} \partial_n$.
Taking equation \eqref{christoffel} into account, the calculation of the curvature tensor and its contractions is straightforward and we get
\begin{equation}
\begin{split}
{R^a}_{bcd} &= \kappa \partial_{[c} {S^a}_{d]b} - \kappa^2 h^a_m \partial_{[c} {S^m}_{d]b} - \frac{1}{2} \kappa^2 \eta_{pm} \eta^{aq} {S^p}_{q[c} {S^m}_{d]b} + \bigo{h^3} \\
R_{ab} &= \frac{1}{2} \kappa \left( \partial_m {S^m}_{ab} - \partial_a \partial_b h \right) - \kappa^2 h^n_m \partial_{[n} {S^m}_{b]a} - \frac{1}{2} \kappa^2 \eta_{mn} \eta^{cd} {S^n}_{d[c} {S^m}_{b]a} + \bigo{h^3} \\
R &= \kappa \left( \partial_m \partial_n h^{mn} - \dalembert h \right) + \kappa^2 h^{mn} \left( \partial_n \partial_m h + \dalembert h_{mn} - 2 \partial_n \partial^a h_{ma} \right) \\
&\quad- \frac{1}{4} \kappa^2 \left( 2 \partial_c h^{nc} - \partial^n h \right) \left( 2 \partial^a h_{na} - \partial_n h \right) + \frac{1}{4} \kappa^2 \left( 3 \partial_c h_{md} - 3 \partial_m h_{cd} \right) \left( \partial^c h^{md} \right) \\
&\quad+ \frac{1}{4} \kappa^2 \left( \partial_d h_{mc} \right) \left( \partial^c h^{md} \right) + \bigo{h^3} \eqend{,}
\end{split}
\end{equation}
with $\dalembert = \eta^{mn} \partial_m \partial_n$.

Finally, the Weyl tensor is given in four dimensions by
\begin{equation}
C_{abcd} = R_{abcd} - \frac{1}{2} \left( R_{ac} g_{bd} - R_{ad} g_{bc} + R_{bd} g_{ac} - R_{bc} g_{ad} \right) + \frac{1}{6} R ( g_{ac} g_{bd} - g_{ad} g_{bc} ) = \bigo{h} \eqend{.}
\end{equation}

\section{Conformal transformation}
\label{appendix_conformal}

Under the conformal transformation
\begin{equation}
\tilde{g}_{ab} = \mathe^{2\omega} g_{ab} \eqend{,}
\end{equation}
where $\omega(x)$ is an arbitrary function of the coordinates, the transformed Christoffel symbols are given by
\begin{equation}
\varchristoffel{a}{b}{c} = \christoffel{a}{b}{c} + \left( \delta^a_b \delta^m_c + \delta^a_c \delta^m_b - g_{bc} g^{am} \right) \partial_m \omega \eqend{.}
\end{equation}
This implies the following transformation for the curvature tensor and its contractions:
\begin{equation}
\begin{split}
{\tilde{R}^a}{}_{bcd} &= {R^a}_{bcd} + 4 g^{ak} \delta^m_{[c} g_{d][k} \left[ \nabla_{b]} \nabla_m \omega - ( \nabla_{b]} \omega ) ( \nabla_m \omega ) \right] - 2 \delta^a_{[c} g_{d]b} ( \nabla^m \omega ) ( \nabla_m \omega ) \\
\tilde{R}_{bd} &= R_{bd} - 2 \left[ \nabla_b \nabla_d \omega - ( \nabla_b \omega ) ( \nabla_d \omega ) + g_{db} ( \nabla^m \omega ) ( \nabla_m \omega ) \right] - g_{bd} \nabla^m \nabla_m \omega \\
\mathe^{2\omega} \tilde{R} &= R - 6 \nabla^m \nabla_m \omega - 6 ( \nabla^m \omega ) ( \nabla_m \omega ) \eqend{,}
\end{split}
\end{equation}
where $\nabla_c$ is the covariant derivative associated with the metric $g_{ab}$.

\section{Special functions}
\label{appendix_special}

We define the entire function $\Ein (z)$ in terms of the following integral:
\begin{equation}
\label{expintegral_ein_definition}
\Ein z = \int_0^z \frac{\mathe^t - 1}{t} \total t = \sum_{k=1}^\infty \frac{z^k}{k \, k!} \eqend{.}
\end{equation}
Its expansion at infinity can be obtained by an asymptotic analysis of the above integral, and is given by
\begin{equation}
\Ein \left( \alpha r + \beta \right) \sim - \gamma - \ln ( - \alpha r ) + \mathe^{\alpha r + \beta} \bigo{r^{-1}} + \bigo{r^{-1}} \qquad (r \to \infty)\eqend{,}
\end{equation}
where $\gamma$ is the Euler-Mascheroni constant, and $\alpha$ and $\beta$ are complex constants with $\Re \alpha \leq 0$.

Some useful integrals involving $\Ein (z)$ which can be easily obtained by partial integration are
\begin{equation}
\label{integrals}
\begin{split}
\int \frac{\mathe^{a t}}{b t + c} \total t &= \frac{1}{b} \mathe^{-\tfrac{a c}{b}} \left[ \Ein\left[ \frac{a}{b} \left( b t + c \right) \right] + \ln \left( b t + c \right) \right] \\
\int \mathe^{a t} \ln ( b t + c ) \total t &= \frac{1}{a} \left[ \left( \mathe^{a t} - \mathe^{-\tfrac{a c}{b}} \right) \ln \left( b t + c \right) - \mathe^{-\tfrac{a c}{b}} \Ein\left[ \frac{a}{b} \left( b t + c \right) \right] \right] \\
\int \mathe^{a t} \Ein ( b t + c ) \total t &= \frac{1}{a} \mathe^{a t} \Ein ( b t + c ) + \frac{1}{a} \mathe^{-\tfrac{a c}{b}} \left[ \Ein\left( \frac{a}{b} \left( b t + c \right) \right) - \Ein\left( \frac{a+b}{b} \left( b t + c \right) \right) \right] \eqend{,}
\end{split}
\end{equation}
where $a$, $b$ and $c$ are arbitrary complex parameters. These integrals depend continuously on the parameters $a$ and $c$. Therefore, the integral of $\Ein (b t + c)$, for instance, can be calculated by taking the limit $a \to 0$ in the last equation of \eqref{integrals}.

\section{Fourier transforms}
\label{appendix_fts}

For the calculation of the Riemann correlator~\eqref{riemann_correlator} we need Fourier transforms of the form
\begin{equation}
f_{i_1 \cdots}(x, x') = \mathcal{F}(\vec{p}, \vec{r}) \left[ \tilde{f}_{i_1 \cdots}(\vec{p}, \eta, \eta') \right] \equiv \int \bar{f}_{i_1 \cdots}(\vec{p}, \eta, \eta')\, \mathe^{\mathi \vec{p} \vec{r}} \frac{\total^3 p}{(2\mathpi)^3} \eqend{,}
\end{equation}
where $\vec{r} = \vec{x} - \vec{x}'$. In order to compute this integrals, we first need to remove the tensorial structure. The tensors $\eta_{ab}$ and $\delta^0_a$ can be simply factored out, whereas $p_a$ factors can be extracted out of the Fourier transform as follows:
\begin{equation}
\label{fouriertrafo_derive}
\mathcal{F}(\vec{p}, \vec{r}) \left[ p_a \bar{f}_{i_1 \cdots}(\vec{p}, \eta, \eta') \right] = - \mathi \left( \partial_a - \delta^0_a \partial_\eta \right) \mathcal{F}(\vec{p}, \vec{r}) \left[ \bar{f}_{i_1 \cdots}(\vec{p}, \eta, \eta') \right] \eqend{.}
\end{equation}
The Fourier transform of the remaining scalar factor must then be understood in the sense of distributions, i.e. we calculate
\begin{equation}
\label{fouriertrafo_distlim}
\begin{split}
f(x, x') &= \distlim_{\epsilon \to 0} \left[ \frac{1}{2 \mathpi^2 \abs{\vec{r}}} \int_0^\infty \mathe^{-\epsilon \abs{\vec{p}}} \tilde{f}(\abs{\vec{p}}, \eta, \eta') \sin\left( \abs{\vec{p}} \abs{\vec{r}} \right) \abs{\vec{p}} \total \abs{\vec{p}} \right] \eqend{,}
\end{split}
\end{equation}
where we have taken into account that because of rotational symmetry those scalar factors only depend on the absolute value of $\vec{p}$.

As a technical detail, we note that for the Fourier transform of e.g.\ $p_a p_b \abs{\vec{p}}^{-3}$ we have to introduce a lower bound $\xi$ for the integral in equation \eqref{fouriertrafo_distlim}. After differentiating as indicated in equation~\eqref{fouriertrafo_derive}, all dependence on $\xi$ disappears. This is because while the Fourier transform of $p_a p_b \abs{\vec{p}}^{-3}$ is well defined (and is given by the result of this procedure), the Fourier transform of $\abs{\vec{p}}^{-3}$ alone is not.

Employing the integrals for the $\Ein$ function given in appendix \ref{appendix_special} and using partial integration to reduce any powers of $\abs{\vec{p}}$, most of the Fourier transforms can be readily calculated in this way. In addition, we need the indefinite integral
\begin{equation}
\int \ln (a p) \frac{1}{p} \total p = \frac{1}{2} \ln^2(a p) \eqend{,}
\end{equation}
and the definite integral
\begin{equation}
I = \int_0^\infty \mathe^{-\epsilon p} \mathe^{\mathi a p} \Ein(\mathi b p) \frac{1}{p} \total p
\end{equation}
for real parameters $a$ and $b$. Since it is absolutely convergent, to compute this integral we can insert the series expansion of the $\Ein$ function to get
\begin{equation}
I = \sum_{k=1}^\infty \frac{(\mathi b)^k}{k \, k!} \int_0^\infty \mathe^{-\epsilon p} \mathe^{\mathi a p} p^{k-1} \total p = \sum_{k=1}^\infty \frac{1}{k^2} \left( - \frac{b}{a + \mathi \epsilon} \right)^k = \Li_2 \left( - \frac{b}{a + \mathi \epsilon} \right) \eqend{,}
\end{equation}
where $\Li_2(z)$ is the dilogarithm function. However, this definite integral only occurs together with $p_a$, so that its derivative,
\begin{equation}
\Li_2'(z) = - \frac{\ln(1-z)}{z} \eqend{,}
\end{equation}
which can also be inferred directly from the series expansion, is the only expression that we need for our computation.

\section{Explicit expressions for the two-point Riemann correlator}
\label{appendix_riemann2pf}

Here we give the explicit expressions for the coefficient functions $\mathcal{R}^{(0,k)}(Z)$ and $\mathcal{R}^{(1,k)}(Z)$ in equation~\eqref{riemann_correlator_pos}. The tree-level results are
\begin{equation}
\begin{aligned}
\mathcal{R}^{(0,1)}(Z) &= \frac{1}{8} (1-3 Z) (1-Z)^{-3} \eqend{,} & \mathcal{R}^{(0,2)}(Z) &= \frac{3}{4} (1-Z)^{-4} \eqend{,} \\
\mathcal{R}^{(0,3)}(Z) &= - \frac{3}{2} (3-Z) (1-Z)^{-5} \eqend{,} & \mathcal{R}^{(0,4)}(Z) &= - \frac{3}{4} (2-Z) (1-Z)^{-4} \eqend{,} \\
\mathcal{R}^{(0,5)}(Z) &= - \frac{3}{4} (1-Z)^{-2} \eqend{,} & \mathcal{R}^{(0,6)}(Z) &= - \frac{3}{4} (1-Z)^{-4} \eqend{,} \\
\mathcal{R}^{(0,7)}(Z) &= \frac{3}{4} (5-Z) (1-Z)^{-5} \eqend{,} & \mathcal{R}^{(0,8)}(Z) &= \frac{1}{4} (4-Z) (1-Z)^{-4} \eqend{,} \\
\mathcal{R}^{(0,9)}(Z) &= \frac{1}{8} (3-Z) (1-Z)^{-3} \eqend{,}
\end{aligned}
\end{equation}
and the matter loop corrections are given by
\begin{equation}
\begin{split}
\mathcal{R}^{(1,1)}(Z) &= - 4 Z^3 (1 + Z)^{-3} (1 - Z)^{-3} \ln\left[ \frac{1}{2} (1 - Z) \right] \\
&\qquad- \frac{1}{6} (1 - 2 Z - Z^2 + 26 Z^3) (1 + Z)^{-2} (1 - Z)^{-4} \eqend{,} \\
\mathcal{R}^{(1,2)}(Z) &= 12 Z (1 + Z^2) (1 + Z)^{-4} (1 - Z)^{-4} \ln\left[ \frac{1}{2} (1 - Z) \right] \\
&\qquad+ Z (11 + 4 Z + 17 Z^2) (1 + Z)^{-3} (1 - Z)^{-5} \eqend{,} \\
\mathcal{R}^{(1,3)}(Z) &= - 12 Z (7 + 10 Z^2 - Z^4) (1 + Z)^{-5} (1 - Z)^{-5} \ln\left[ \frac{1}{2} (1 - Z) \right] \\
&\qquad- (8 + 101 Z + 58 Z^2 + 168 Z^3 - 2 Z^4 - 13 Z^5) (1 + Z)^{-4} (1 - Z)^{-6} \eqend{,} \\
\mathcal{R}^{(1,4)}(Z) &= - 6 (1 + 4 Z^2 - Z^4) (1 + Z)^{-4} (1 - Z)^{-4} \ln\left[ \frac{1}{2} (1 - Z) \right] \\
&\qquad- \frac{1}{2} (11 + 10 Z + 58 Z^2 - 2 Z^3 - 13 Z^4) (1 + Z)^{-3} (1 - Z)^{-5} \eqend{,} \\
\mathcal{R}^{(1,5)}(Z) &= - 6 Z (1 + Z)^{-2} (1 - Z)^{-2} \ln\left[ \frac{1}{2} (1 - Z) \right] - \frac{1}{2} Z (11 - 13 Z) (1 + Z)^{-1} (1 - Z)^{-4} \eqend{,} \\
\mathcal{R}^{(1,6)}(Z) &= - 12 Z (1 + Z^2) (1 + Z)^{-4} (1 - Z)^{-4} \ln\left[ \frac{1}{2} (1 - Z) \right] \\
&\qquad- (2 + 17 Z + 10 Z^2 + 19 Z^3) (1 + Z)^{-3} (1 - Z)^{-5} \eqend{,} \\
\mathcal{R}^{(1,7)}(Z) &= 24 Z (3 + 5 Z^2) (1 + Z)^{-5} (1 - Z)^{-5} \ln\left[ \frac{1}{2} (1 - Z) \right] \\
&\qquad+ 2 (11 + 71 Z + 61 Z^2 + 97 Z^3) (1 + Z)^{-4} (1 - Z)^{-6} \eqend{,} \\
\mathcal{R}^{(1,8)}(Z) &= 4 (1 + 5 Z^2) (1 + Z)^{-4} (1 - Z)^{-4} \ln\left[ \frac{1}{2} (1 - Z) \right] \\
&\qquad+ (7 + 12 Z + 29 Z^2) (1 + Z)^{-3} (1 - Z)^{-5} \eqend{,} \\
\mathcal{R}^{(1,9)}(Z) &= 4 Z (1 + Z)^{-3} (1 - Z)^{-3} \ln\left[ \frac{1}{2} (1 - Z) \right] + (1 + 5 Z) (1 + Z)^{-2} (1 - Z)^{-4} \eqend{.}
\end{split}
\end{equation}

Using the basis introduced by Allen and Jacobson \cite{allenjacobson86}, the tree-level coefficient functions $\mathcal{S}^{(0,k)}(Z)$ in equation~\eqref{riemann_correlator_nn} are given by
\begin{equation}
\begin{aligned}
\mathcal{S}^{(0,1)}(Z) &= \frac{1}{8} (1-3 Z) (1-Z)^{-3} \eqend{,} & \mathcal{S}^{(0,2)}(Z) &= \frac{3}{4} (1+Z) (1-Z)^{-3} \eqend{,} \\
\mathcal{S}^{(0,3)}(Z) &= - 12 (1-Z)^{-3} \eqend{,} & \mathcal{S}^{(0,4)}(Z) &= - \frac{3}{4} (3-Z) (1-Z)^{-3} \eqend{,} \\
\mathcal{S}^{(0,5)}(Z) &= - \frac{3}{4} (1-Z)^{-2} \eqend{,} & \mathcal{S}^{(0,6)}(Z) &= - \frac{3}{4} (1+Z) (1-Z)^{-3} \eqend{,} \\
\mathcal{S}^{(0,7)}(Z) &= 12 (1-Z)^{-3} \eqend{,} & \mathcal{S}^{(0,8)}(Z) &= \frac{1}{4} (7-Z) (1-Z)^{-3} \eqend{,} \\
\mathcal{S}^{(0,9)}(Z) &= \frac{1}{8} (3-Z) (1-Z)^{-3} \eqend{,}
\end{aligned}
\end{equation}
and the matter loop corrections $\mathcal{S}^{(1,k)}$ are
\begin{equation}
\begin{split}
\mathcal{S}^{(1,1)}(Z) &= - 4 Z^3 (1 + Z)^{-3} (1 - Z)^{-3} \ln\left[ \frac{1}{2} (1 - Z) \right] \\
&\qquad- \frac{1}{6} (1 - 2 Z - Z^2 + 26 Z^3) (1 + Z)^{-2} (1 - Z)^{-4} \eqend{,} \\
\mathcal{S}^{(1,2)}(Z) &= 12 Z (1 + Z^2) (1 + Z)^{-3} (1 - Z)^{-3} \ln\left[ \frac{1}{2} (1 - Z) \right] \\
&\qquad+ Z (11 + 4 Z + 17 Z^2) (1 + Z)^{-2} (1 - Z)^{-4} \eqend{,} \\
\mathcal{S}^{(1,3)}(Z) &= - 24 (1 - Z)^{-3} \ln\left[ \frac{1}{2} (1 - Z) \right] - 10 (3 + 5 Z) (1 - Z)^{-4} \eqend{,} \\
\mathcal{S}^{(1,4)}(Z) &= - 6 (1 + Z + 3 Z^2 - Z^3) (1 + Z)^{-3} (1 - Z)^{-3} \ln\left[ \frac{1}{2} (1 - Z) \right] \\
&\qquad- \frac{1}{2} (11 + 21 Z + 45 Z^2 - 13 Z^3 ) (1 + Z)^{-2} (1 - Z)^{-4} \eqend{,} \\
\mathcal{S}^{(1,5)}(Z) &= - 6 Z (1 + Z)^{-2} (1 - Z)^{-2} \ln\left[ \frac{1}{2} (1 - Z) \right] - \frac{1}{2} Z (11 - 13 Z) (1 + Z)^{-1} (1 - Z)^{-4} \eqend{,} \\
\mathcal{S}^{(1,6)}(Z) &= - 12 Z (1 + Z^2) (1 + Z)^{-3} (1 - Z)^{-3} \ln\left[ \frac{1}{2} (1 - Z) \right] \\
&\qquad- (2 + 17 Z + 10 Z^2 + 19 Z^3) (1 + Z)^{-2} (1 - Z)^{-4} \eqend{,} \\
\mathcal{S}^{(1,7)}(Z) &= 24 (1 - Z)^{-3} \ln\left[ \frac{1}{2} (1 - Z) \right] + 10 (7 + 5 Z) (1 - Z)^{-4} \eqend{,} \\
\mathcal{S}^{(1,8)}(Z) &= 4 (1 + 2 Z + 3 Z^2) (1 + Z)^{-3} (1 - Z)^{-3} \ln\left[ \frac{1}{2} (1 - Z) \right] \\
&\qquad+ (9 + 20 Z + 19 Z^2) (1 + Z)^{-2} (1 - Z)^{-4} \eqend{,} \\
\mathcal{S}^{(1,9)}(Z) &= 4 Z (1 + Z)^{-3} (1 - Z)^{-3} \ln\left[ \frac{1}{2} (1 - Z) \right] + (1 + 5 Z) (1 + Z)^{-2} (1 - Z)^{-4} \eqend{.}
\end{split}
\end{equation}

\bibliography{literature}

\end{document}